\documentclass[final,5p,times,twocolumn]{elsarticle}

\usepackage{amssymb}
\usepackage{amsmath}
\usepackage[dvipsnames]{xcolor}
\usepackage{placeins}
\usepackage{hyperref}
\hypersetup{hidelinks}
\usepackage{microtype}
\usepackage{booktabs}
\usepackage{multirow}
\usepackage{makecell}

\usepackage{url}

\newcommand{\coordX}{x}
\newcommand{\coordY}{y}
\newcommand{\coordZ}{z}

\newcommand{\stateMatrix}{\boldsymbol{U}^{\prime}}
\newcommand{\stateMatrixRec}{\boldsymbol{\tilde{U}}^{\prime}}
\newcommand{\stateMatrixTime}[1]{\boldsymbol{U}^{\prime(#1)}}
\newcommand{\stateMatrixRecTime}[1]{\boldsymbol{\tilde{U}}^{\prime(#1)}}

\newcommand{\numberOfGridPoints}{N_{p}}
\newcommand{\numberOfGridPointsIdx}[1]{N_{p_{#1}}}

\newcommand{\PODModalAmpVecTime}[1]{\boldsymbol{a}^{(#1)}}

\newcommand{\latentVariableVecTime}[1]{\boldsymbol{z}_{\mathrm{lat}}^{(#1)}}
\newcommand{\latentVariableVecIdx}[1]{\boldsymbol{z}_{\mathrm{lat}, \, #1}}
\newcommand{\latentVariableVec}{\boldsymbol{z}_{\mathrm{lat}}}

\newcommand{\parameterSetEncoder}{\boldsymbol{\Psi}}
\newcommand{\parameterSetDecoder}{\boldsymbol{\Omega}}

\DeclareMathOperator*{\argmin}{argmin}

\journal{Computers \& Fluids}

\begin{document}

\begin{frontmatter}

\title{A Priori Estimation of the Non-Linear Compression Potential and Quantification of Autoencoder Non-Linearity in Turbulent Flows}

\author[1]{Niklas Alexander Pissarski\texorpdfstring{\corref{cor1}}{}}
\ead{nalpi@dtu.dk}
\author[2]{Juan Pablo Murcia Leon}
\author[1]{Søren Juhl Andersen}

\cortext[cor1]{Corresponding author}

\affiliation[1]{organization={Department of Wind and Energy Systems, Technical University of Denmark},
postcode={2800},
city={Kgs. Lyngby},
country={Denmark}}
\affiliation[2]{organization={Department of Wind and Energy Systems, Technical University of Denmark},
postcode={4000},
city={Roskilde},
country={Denmark}}

\begin{abstract}
Data-driven reduced-order modelling (ROM) of turbulent flows is based on an efficient and physical low-dimensional representation of the flow process. Such a dimension reduction can be achieved using proper orthogonal decomposition (POD) or non-linear autoencoders. Capturing the inherent non-linearities in turbulent flows offers the potential for improved compression rates. This work proposes the local tangent residual (LTR) as an a priori estimate of the potential benefit of non-linear dimension reduction for a given dataset. The LTR is based on the local geometry of the data points, requiring neither prior model training nor the definition of a threshold parameter. For all datasets investigated in this work, namely, two vortex street cases and the turbulent wake flow of a wind turbine, the dominant trends in autoencoder compression performance are reproduced by the proposed LTR. The results are further contextualised through the decoder linearisation error (DLE). The DLE quantifies the non-linearity of the trained decoder by comparing it to a linearised surrogate model, which enables physical interpretability of the non-linearities. Together, a large non-linear compression potential in the turbulent flow, as quantified by the LTR, is corroborated by a large DLE, i.e., a pronounced non-linearity captured in the trained autoencoder. The LTR thus enables practitioners in turbulent flow modelling to determine a priori whether non-linear compression is worthwhile, avoiding the substantial computational cost of autoencoder architecture search and hyperparameter tuning in cases where POD already suffices.
\end{abstract}


\end{frontmatter}

\section{Introduction}
\label{sec:introduction}

Modelling turbulence remains one of the most fundamental and important challenges in classical physics, as turbulent flows govern phenomena that range from global climate processes to the combustion in engines and the pumping of the human heart. The difficulties arise from the non-linear nature of turbulence: although numerical approximations are possible at a high computational cost, chaotic behaviour means that the solutions are sensitive to small changes in the parameters (the \emph{butterfly effect}). Moreover, these studies require the use of substantial resources~\cite{Yang2024}. Therefore, physical insights are often limited to specific cases.

The increasing prominence of data-driven \emph{reduced-order modelling} (ROM) holds the promise of a fundamental paradigm shift: fast and accurate data-driven modelling can be applied to predict specific flow realisations in unseen conditions~\cite{Meneveau2019}. ROM is a substantial challenge due to the inherent characteristics of turbulence, which is a high-dimensional phenomenon, spanning a wide range of energetically significant spatial and temporal scales with non-linear interactions~\cite{Meneveau2019}.

\emph{Dimension reduction methods} include linear decomposition methods such as \emph{proper orthogonal decomposition} (POD)~\cite{Lumley1967, Sirovich1987, Berkooz1993} or nonlinear neural network-based approaches such as \emph{convolutional autoencoders} (CAE)~\cite{Milano2002, Masci2011, Morimoto2021}. The standard ROM pipeline in turbulent flow modelling includes projecting high-resolution flow snapshots (i.e., samples of the velocity field at different times) onto a low-dimensional latent space~\cite{Aubry1988}. This procedure constitutes the foundation for modelling frameworks like the \emph{predictive-stochastic} PS-ROM~\cite{Andersen2022}.

Several studies have empirically demonstrated that non-linear compression can be more efficient than POD~\cite{Murata2019, Eivazi2022, Fukami2020, Doan2023, SoleraRico2024a}. This advantage is particularly pronounced for flows exhibiting periodic vortex shedding~\cite{Murata2019, Eivazi2022, Fukami2020, SoleraRico2024a}, whereas for turbulent flows containing a broad range of scales, such as turbulent channel flows, the performance gains of a conventional autoencoder over POD have been found to be marginal~\cite{Fukami2020, Fukami2021}.

At the same time, training an autoencoder requires substantial computational effort, a large number of architectural decisions, and extensive hyperparameter tuning, while POD is parameter-free and efficiently computed via sparse \emph{singular value decomposition} (SVD). This raises the practical question of whether the additional complexities of autoencoders are warranted for a given dataset and how this can be assessed without training the network first. It is therefore worth examining which characteristic properties of a turbulent flow lead to substantial performance gains over POD and under which conditions the advantage of using an autoencoder is marginal or absent altogether.

In a natural manner, the \emph{intrinsic dimension} of the dataset characterises whether it is suitable for dimension reduction, where a lower intrinsic dimension indicates that the data can be represented using fewer latent coordinates without significant loss of information~\cite{Binnie2025}. Existing approaches to estimate the intrinsic dimension include \emph{local}-PCA~\cite{Fan2010, Fukunaga1971, Lin2008} methods, in which the number of dominant eigenvalues of a PCA computed within a small neighbourhood of data points serves as an estimate of the local manifold dimension. The central difficulty in local-PCA methods lies in the selection of a threshold value that defines what constitutes a dominant eigenvalue; different strategies have been proposed in this context~\cite{Kaiser1960, Jolliffe2004}. A comprehensive overview of dimension estimation methods, and in particular the \emph{tangential estimators} including local-PCA, is given in~\cite{Binnie2025}.

The central contribution of this work is the introduction of two metrics: (1) the \emph{local tangent residual} (LTR) and (2) the \emph{decoder linearisation error} (DLE). The LTR provides an a priori estimate of the potential CAE benefit over POD for a given dataset, prior to any computationally demanding neural network training or architectural decisions and without requiring the definition of a threshold parameter. To empirically validate this metric, a systematic comparison between CAE and POD is conducted for three flow cases: a wind turbine wake and two reference cases, namely, the flow around two collinear flat plates at different \textsc{Reynolds} numbers that exhibits periodic vortex shedding. Here, POD serves as an optimal linear baseline. The DLE contextualises these results by quantifying the non-linearity of the trained decoder using a linearised surrogate model.

The remainder of this paper is organised as follows: Section~\ref{sec:methods:dimension_reduction} introduces the considered dimension reduction methods along with the applied neural network architecture and training procedure. The two proposed metrics (LTR and DLE) are presented in Section~\ref{sec:methods:metrics} and the validation datasets are detailed in Section~\ref{sec:methods:datasets}. Section~\ref{sec:results} presents and compares the reconstruction performance of both models alongside the metric results, which are then discussed in Section~\ref{sec:discussion}.

\section{Methods}
\label{sec:methods}

This work examines ROMs that describe the temporal evolution of the turbulent flow using the fluctuating component from the \textsc{Reynolds} decomposition: 

\begin{equation}
    \stateMatrix(\boldsymbol{X}, \, t) = \boldsymbol{U}(\boldsymbol{X}, \, t) - \overline{\boldsymbol{U}}(\boldsymbol{X}) \, ,
\end{equation}

\noindent where $ \boldsymbol{U}(\boldsymbol{X}, \, t) = \bigl[u(\boldsymbol{X}, \, t) \quad v(\boldsymbol{X}, \, t) \quad w(\boldsymbol{X}, \, t)\bigr]^{\top}$ is the velocity vector at position $\boldsymbol{X} = \bigl[\coordX \quad \coordY\bigr]$ and $\overline{\boldsymbol{U}}(\boldsymbol{X})$ denotes the time-averaged velocity.

In the context of dimension reduction, it is standard practice to form a snapshot matrix $\stateMatrix \in \mathbb{R}^{3\numberOfGridPoints \times T}$ from $T$ independent flow field observations. Each of the $T$ columns of this matrix represents a single high-dimensional snapshot $\stateMatrixTime{t} \in \mathbb{R}^{3\numberOfGridPoints}$ containing the three fluctuating velocity components $\boldsymbol{u}^{\prime}, \, \boldsymbol{v}^{\prime}, \, \boldsymbol{w}^{\prime} \in \mathbb{R}^{\numberOfGridPoints}$ at $\numberOfGridPoints$ coordinate points for timestep $t$.

\subsection{Dimension Reduction}
\label{sec:methods:dimension_reduction}

Two methods are considered for dimension reduction of the flow: POD as the baseline linear dimension reduction method and the CAE for the non-linear generalisation of this concept.

\subsubsection{Proper orthogonal decomposition (POD)}
\label{sec:methods:dimension_reduction:pod}

POD is a linear decomposition method for turbulent flows which was introduced by \textsc{Lumley}~\cite{Lumley1967}. It represents the flow field fluctuations $\stateMatrix$ as a linear superposition of static spatial patterns $\boldsymbol{\Phi}_{i} \in \mathbb{R}^{3\numberOfGridPoints}$ that are referred to as \emph{modes} and corresponding temporal coefficients $a_{i}^{(t)} \in \mathbb{R}$ that are termed \emph{modal amplitudes}:

\begin{equation}
    \boldsymbol{U}^{(t)} = \overline{\boldsymbol{U}} + \sum_{i=1}^{3\numberOfGridPoints} \, a_{i}^{(t)}\boldsymbol{\Phi}_{i} \, .
\end{equation}

\noindent In this representation, the modes serve as an \emph{orthonormal} basis for the fluctuating flow field snapshots. POD defines this basis such that each subsequent mode captures the maximum possible remaining \emph{turbulent kinetic energy} (TKE). Consequently, the TKE in each modal amplitude is maximised in descending order, allowing the modes to extract the most energetic spatial structures in a hierarchical manner.

The full \emph{orthonormal} basis consists of $3\numberOfGridPoints$ POD modes $\boldsymbol{\Phi} \in \mathbb{R}^{3\numberOfGridPoints \times 3\numberOfGridPoints}$ that correspond to the principal axes $\boldsymbol{V}$ of the covariance matrix $\boldsymbol{C} = \frac{1}{T} \, \stateMatrix{\stateMatrix}^{\top} \in \mathbb{R}^{3\numberOfGridPoints \times 3\numberOfGridPoints}$. The principal axes are determined in solving the eigenvalue problem: $\boldsymbol{C} = \boldsymbol{V}\boldsymbol{\Lambda}\boldsymbol{V}^{\top}$ with $\boldsymbol{V}^{-1} = \boldsymbol{V}^{\top}$.

In the frequent case where $3\numberOfGridPoints > T$, the more efficient \emph{snapshot method}~\cite{Sirovich1987} can be applied. Here, the eigendecomposition is performed on the smaller temporal correlation matrix $\boldsymbol{C}_{\mathrm{snap}} = \frac{1}{T} \, {\stateMatrix}^{\top}\stateMatrix \in \mathbb{R}^{T \times T}$. Let $\boldsymbol{C}_{\mathrm{snap}} = \boldsymbol{V}_{\mathrm{snap}}\boldsymbol{\Lambda}{\boldsymbol{V}_{\mathrm{snap}}}^{\top}$ be the resulting decomposition, where $\boldsymbol{V}_{\mathrm{snap}}$ contains the temporal eigenvectors and $\boldsymbol{\Lambda}$ is the diagonal matrix of eigenvalues. The spatial POD modes are then recovered via: $\boldsymbol{\Phi} = \stateMatrix\boldsymbol{V}_{\mathrm{snap}}{(T\boldsymbol{\Lambda})}^{-1/2}$.

Since the POD modes are associated with eigenvalues $\lambda_{i}$ which are sorted in descending order and represent the TKE captured in each mode, a reduced-order representation of the flow can be reconstructed using the following:

\begin{equation}
    \boldsymbol{\tilde{U}}^{(t)} = \overline{\boldsymbol{U}} + \sum_{i=1}^{d_{\mathrm{lat}}} \, a_{i}^{(t)}\boldsymbol{\Phi}_{i} \, .
\end{equation}

\begin{figure}[tb]
    \centering
    \includegraphics{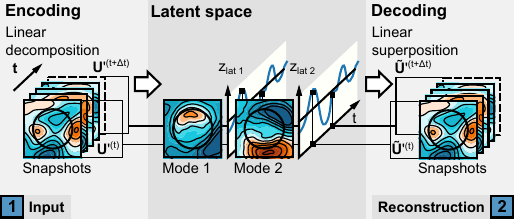}
    \caption{\textbf{Linear dimension reduction using POD with two modes.} The modal time series constitute the latent space of the POD representation.}
    \label{fig:PS-ROM_POD_small}
\end{figure}

The \emph{reconstructed} snapshot $\boldsymbol{\tilde{U}}^{(t)}$, or equivalently its fluctuating part $\stateMatrixRecTime{t}$, is a reduced-order approximation of the original input snapshot $\stateMatrixTime{t}$. For this truncated representation, not the full set of $3\numberOfGridPoints$ POD modes is used but the first $d_{\mathrm{lat}}$ modes. This truncated set of modes spans a linear subspace with $d_{\mathrm{lat}}$ dimensions, also termed \emph{latent} space here. Hence, the modal amplitude $\PODModalAmpVecTime{t}$ is also referred to as latent coordinate $\latentVariableVecTime{t}$ in the following, as shown in Figure~\ref{fig:PS-ROM_POD_small}.

\subsubsection{Convolutional autoencoder (CAE)}
\label{sec:methods:dimension_reduction:cae}

\begin{figure}[tb]
    \centering
    \includegraphics{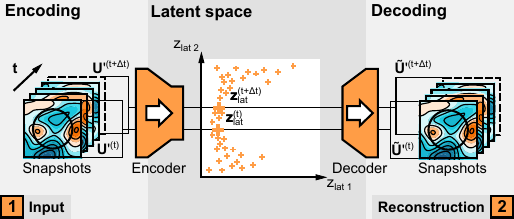}
    \caption{\textbf{Non-linear dimension reduction using an autoencoder with two latent dimensions.} The autoencoder is presented as an encoder--decoder pair.}
    \label{fig:PS-ROM_CAE_small}
\end{figure}

An autoencoder is a neural network that uses a parameterised encoder to convert snapshots $\stateMatrixTime{t} \in \mathbb{R}^{3\numberOfGridPoints}$ into a low-dimensional latent representation:

\begin{equation}
    \label{eqn:encoder}
    \latentVariableVecTime{t} = f_{\parameterSetEncoder}(\stateMatrixTime{t}) \quad \text{with} \quad f_{\parameterSetEncoder} \in \mathcal{F} \, ,
\end{equation}

\noindent where $f_{\parameterSetEncoder}(\cdot)$ represents the non-linear mapping function with parameters $\parameterSetEncoder$, and $\mathcal{F}$ the space of possible mappings from the high-dimensional data space to the latent space. The latent representation $\latentVariableVecTime{t}$ encodes the input snapshot, while the output snapshot is reconstructed from the low-dimensional representation in the decoder:

\begin{equation}
    \label{eqn:decoder}
    \stateMatrixRecTime{t} = g_{\parameterSetDecoder}(\latentVariableVecTime{t}) \quad \text{with} \quad g_{\parameterSetDecoder} \in \mathcal{G} \, ,
\end{equation}

\noindent where $g_{\parameterSetDecoder}(\cdot)$ represents the non-linear mapping function with parameters $\parameterSetDecoder$, and $\mathcal{G}$ the space of possible mappings from the latent space back to the high-dimensional data space. This concept of encoding and decoding snapshots in an autoencoder is illustrated in Figure~\ref{fig:PS-ROM_CAE_small}.

Prior to its application for dimension reduction, the autoencoder must first be  ``trained'' for a minimised reconstruction error between input and output snapshots. In the training process, the trainable parameters $\parameterSetEncoder$ and $\parameterSetDecoder$ that govern the respective mapping functions are varied using gradient-based optimisation methods.

To represent these mapping functions, an autoencoder consists of multiple neuron layers, including at least an input layer, a number of \emph{hidden} layers, and an output layer. For reconstructing the input data, the input and output layers have the same number of neurons ($3\numberOfGridPoints$ in this case for the three fluctuating components) representing the input and output snapshots. The bottleneck layer with the fewest number of neurons $d_{\mathrm{lat}}$ stores the lower-dimensional latent representation of the snapshot. For a general network, the neuron values for a given \emph{dense layer} ($l$) are then calculated based on each individual neuron value in the preceding layer ($l-1$).

\begin{equation}
    \label{eqn:layer_dense}
    \boldsymbol{h}^{[l]} = \varphi(\boldsymbol{W}^{[l]}\boldsymbol{h}^{[l-1]} + \boldsymbol{b}^{[l]}) \, ,
\end{equation}

\noindent where $\boldsymbol{h}^{[l]} \in \mathbb{R}^{N_{l}}$ denotes the output vector and $\boldsymbol{h}^{[l-1]} \in \mathbb{R}^{N_{l-1}}$ the input vector (output of the preceding \emph{hidden} layer) with $\boldsymbol{h}^{[0]} = \stateMatrixTime{t}$. The weight matrix is denoted as $\boldsymbol{W}^{[l]} \in \mathbb{R}^{N_{l} \times N_{l-1}}$ and serves to compute a weighted sum of the input neuron values for \emph{each} output neuron considering a bias $\boldsymbol{b} \in \mathbb{R}^{N_{l}}$. The individual elements of the weight matrices and bias vectors throughout the network form the trainable parameters $\parameterSetEncoder$ and $\parameterSetDecoder$ that are optimised during training.

The function $\varphi(\cdot)$ references the activation function, which is applied elementwise. As long as the activation function is linear, the encoder--decoder mapping and thus the autoencoder as a whole are linear. The choice of activation function can hence introduce non-linearities into the network to increase compression rates compared to linear dimension reduction.

CAEs~\cite{Masci2011} are a variant of autoencoders that take advantage of the spatio-coherent characteristics of turbulent flows using \emph{convolutional layers}~\cite{LeCun1989} (see \ref{app:mathematical_formulation:convolutional_layer} for a formal definition) instead of \emph{dense layers}. These are based on the fundamental concepts of both local receptive fields and shared weights. A local receptive field means that each neuron processes information from a local region of the input snapshot. The concept of shared weights, in contrast, ensures that the same set of weights is used across different regions of the input snapshot, which reduces the number of trainable parameters. These concepts are realised through a trainable filter kernel that convolves the input snapshot to produce a so-called \emph{feature map}. Multiple filter kernels produce multiple feature maps that serve as input to the next convolutional layer and thus enable the network to detect different spatial patterns. 

\subsubsection{Comparing POD and CAE}
\label{sec:methods:dimension_reduction:pod_versus_cae}

Under certain conditions, POD constitutes the optimal linear baseline for an autoencoder. In this interpretation, the linear encoder and decoder are defined using the POD modes with:

\begin{align}
    \label{eqn:encoder_POD}
    \latentVariableVecTime{t} &= \boldsymbol{\Phi}^{\top}\stateMatrixTime{t} \quad \text{(encoder)} ,\\
    \label{eqn:decoder_POD}
    \stateMatrixRecTime{t} &= \boldsymbol{\Phi}\latentVariableVecTime{t} \quad \text{(decoder)} .
\end{align}

\noindent It can be shown that POD and linear autoencoders span the same linear subspace: provided that the autoencoder is single-layer, linear, and bias-free, and is trained on minimising the \emph{mean-squared error} (MSE), then the global MSE-minimum coincides with the POD subspace~\cite{Baldi1989}.

Following this argumentation, the MSE is the natural choice for comparisons between autoencoders and POD, as POD only constitutes the optimal linear baseline method for exactly this error definition. In this work, the \emph{normalised} form of the MSE (NMSE) is thus used to quantify the reconstruction error between the reconstructed snapshot $\stateMatrixRecTime{t}$ and the input snapshot $\stateMatrixTime{t}$:

\begin{equation}
    \label{eqn:nmse}
	e_{\mathrm{NMSE}}\bigl(\stateMatrixRecTime{t}, \, \stateMatrixTime{t}\bigr) = \frac{ \sum_{t \in \mathcal{T}} \bigl\|\stateMatrixRecTime{t} - \stateMatrixTime{t}\bigr\|_{2}^{2}}{\sum_{t \in \mathcal{T}} \bigl\|\stateMatrixTime{t}\bigr\|_{2}^{2}} \, ,
\end{equation}

\noindent where the difference in terms of TKE is normalised with the dataset TKE. Hence, the NMSE is proportional to the amount of TKE that is missing for complete reconstruction.

\subsubsection{CAE architecture search}
\label{sec:methods:dimension_reduction:cae_architecture_search}

\begin{figure*}[tb]
    \centering
    \includegraphics{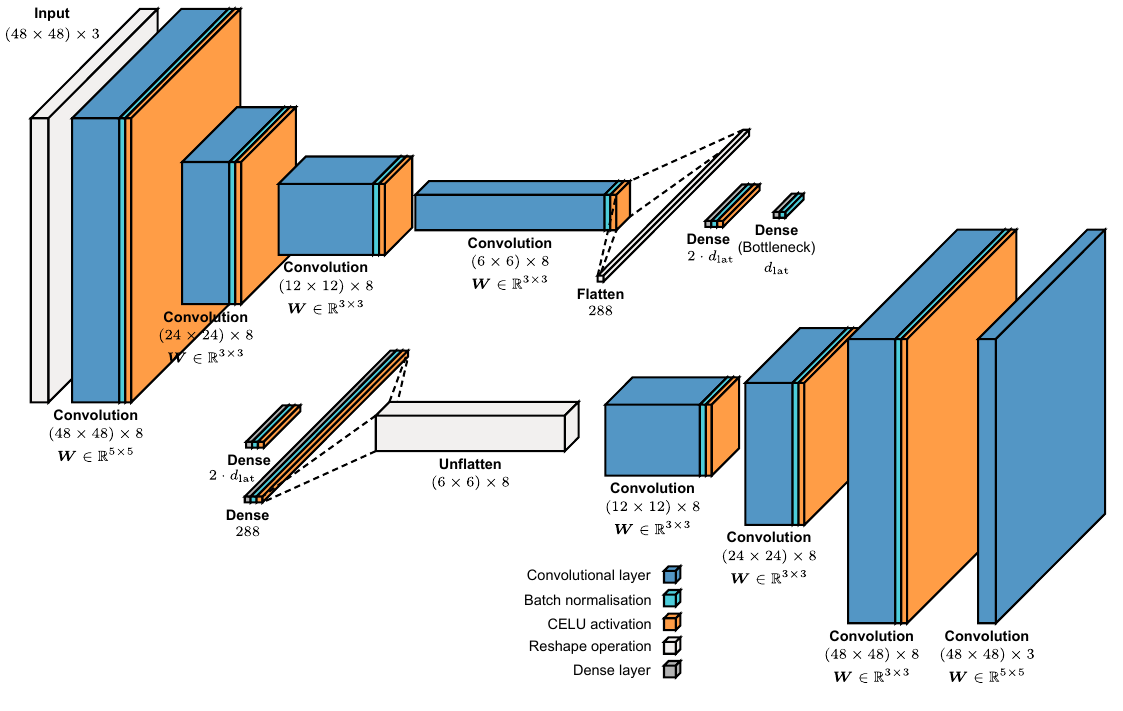} 
    \caption{\textbf{Diagram of the CAE architecture used.} All boxes indicate the dimensions of the output of the colour-coded operation. The convolutional units are represented using boxes, where the height and depth of each box represent the spatial dimensions and the width of the box represents the number of feature maps. The dimensions of feature maps are given in parentheses. The figure is not to scale.}
    \label{fig:autoencoder_architecture}
\end{figure*}

The CAE architecture applied in this work is illustrated in Figure~\ref{fig:autoencoder_architecture}. Both the encoder and decoder comprise four stages, each of which consists of a convolutional layer with either downsampling in the encoder case or upsampling in the decoder case, a subsequent \emph{batch normalisation}\footnote{Batch normalisation is applied to stabilise the training process through preventing \emph{internal covariate shift}, i.e., the significant change in the distribution of inputs from layer to layer~\cite{Ioffe2015}.}~\cite{Ioffe2015} and an activation function.

In the encoder, the downsampling results in a reduction in the input snapshots' spatial resolution that is achieved through strided convolutions where the filter kernel skips a defined number of grid points $n_{\mathrm{skip}}$. In this case, the first stage uses a standard convolution keeping the spatial dimensions constant ($n^{[0]}_{\mathrm{skip}} = 0$) while the subsequent stages reduce the input dimensions with a factor of two ($n^{[l]}_{\mathrm{skip}} = 1$). This architecture downsamples the spatial resolution from $48 \times 48$ to $6 \times 6$ grid points, while the encoder expands the three input components to eight feature maps.

The output of the last encoding stage is flattened to a vector and further reduced using a dense subnetwork consisting of a dense layer with $2 \cdot d_{\mathrm{lat}}$ neurons and another dense layer that produces the $d_{\mathrm{lat}}$-dimensional bottleneck vector with subsequent batch normalisation without trainable parameters that standardises the latent variable $\latentVariableVecTime{t}$ to have zero mean and unit variance\footnote{The mean and standard deviation are estimated based on batches, so the zero mean and unit variance specifications are approximate.}. No activation function is applied to the bottleneck vector, i.e., the mapping is linear.

In the decoder, the upsampling is performed using bilinear upsampling\footnote{Bilinear upsampling is an interpolation technique that increases the spatial resolution through the computation of output pixel values as weighted averages of the four nearest input pixels.} followed by a standard convolution ($n^{[0]}_{\mathrm{skip}} = 0$) to provide trainable parameters. Mirroring the encoder, the final stage in the decoder omits further upsampling. Furthermore, no activation function or normalisation is applied so as not to limit the output.

The initial weights $\boldsymbol{W}$ of the convolutional layers throughout the neural network are generated using a \textsc{Kaiming} normal distribution~\cite{He2015}, and the initial weights of the dense subnetwork are generated with matrix orthogonalisation~\cite{Saxe2013}. Here, the \textsc{Kaiming} method enforces weight scaling depending on the number of ingoing and outgoing neuron connections to stabilise the training of deep neural networks.

The proposed CAE architecture is based on the \emph{continuously differentiable exponential linear unit} (CELU)~\cite{Barron2017} activation function that is defined as

\begin{equation}
	\mathrm{CELU}(x, \, \alpha) =
	\begin{cases}
		x & \text{if } x \geq 0\\
		\alpha (e^{\frac{x}{\alpha}} - 1) & \text{otherwise}
	\end{cases}
\end{equation}

In contrast to the related \emph{rectified linear unit} (ReLU) activation function with $\mathrm{ReLU}(x) = \max(0, \, x)$, the $\mathrm{CELU}$ is continuously differentiable~\cite{Barron2017}. The CELU parameter ($\alpha$) can be interpreted as a ``linearisation'' parameter for the activation function; for $\alpha \to \infty$, $\mathrm{CELU}$ approaches the linear function $\varphi(x) = x$, whereas smaller values increase the non-linearity in the negative input range. The parameter is set to $\alpha = 5$.

The proposed CAE architecture and training configuration were chosen based on the results of two hyperparameter studies covering $100$ different architectural and training configurations each. The tuned hyperparameters are: the size of the dense subnetwork in the bottleneck (i.e., the number of additional dense layers), the number of feature maps throughout the network, the input and output kernel size, the $\alpha$ parameter and the learning rate and batch size. The studies were based on the \emph{tree-structured} \textsc{Parzen} \emph{estimator} (TPE) algorithm (implemented in Optuna~\cite{Akiba2019}, version 4.1.0). The details of the hyperparameter studies are presented in~\ref{app:hyperparameter_optimisation}. The CAE is implemented in PyTorch~\cite{Paszke2019} (v.2.5.1). Table~\ref{tab:trainable_parameters} provides an overview of the number of trainable parameters for the smallest, the selected and the largest considered CAE configuration in the hyperparameter studies and Table~\ref{tab:hyperparameters} presents the corresponding hyperparameters of the CAE architecture search.

\begin{table}[htb]
    \centering
    \footnotesize
    \caption{\textbf{Total number of trainable parameters.} List of parameter numbers depending on the number of latent dimensions for the smallest, the selected and the largest considered CAE configuration.}
    \label{tab:trainable_parameters}
    
    \begin{tabular}{ l l l l l l l l l }
        \toprule
        \multicolumn{9}{c}{$n_{\text{params}}$ [$\times 10^3$]} \\
        \midrule
        $d_{\text{lat}}$ & 1 & 2 & 4 & 8 & 16 & 32 & 64 & 128 \\
        \midrule
        smallest & 5 & 6 & 8 & 12 & 18 & 32 & 60 & 115 \\
        selected & 7 & 8 & 10 & 15 & 26 & 49 & 104 & 252 \\
        largest & 27 & 36 & 55 & 93 & 171 & 336 & 699 & $1 \, 561$ \\
        \bottomrule
    \end{tabular}
\end{table}

\begin{table*}[htb]
    \centering
    \small
    \setlength{\tabcolsep}{4.5pt}
    \newcommand{\desc}[1]{\footnotesize\color{black!70}#1}
    \caption{\textbf{Hyperparameters of the CAE architecture search.} For each discrete hyperparameter, all options of the search space are listed. (\checkmark) marks the option used in the selected configuration and in the networks with the fewest and the most trainable parameters that the search space permits.}
    \label{tab:hyperparameters}
    \begin{tabular}{@{}lllccc@{}}
        \toprule
        Hyperparameter & Option & Definition & Selected network                                 & Smallest network & Largest network \\
                       &        &            & \desc{(Figure~\ref{fig:autoencoder_architecture})} & \desc{(fewest parameters,} & \desc{(most parameters,} \\
                       &        &            &                                                  & \desc{Figure~\ref{fig:Autoencoder_small})} & \desc{Figure~\ref{fig:Autoencoder_large})} \\
        \midrule
        \multirow{3}{*}{\makecell[l]{Additional dense layers\\ \quad \desc{neurons per dense layer in the}\\ \quad \desc{encoder after the ``flatten'' operation}}}
             & 0    & ``flatten'' $\to d_{\mathrm{lat}}$                 &            & \checkmark &            \\
             & 1    & ``flatten'' $\to 2 \cdot d_{\mathrm{lat}} \to d_{\mathrm{lat}}$          & \checkmark &            &            \\
             & 2    & ``flatten'' $\to 4 \cdot d_{\mathrm{lat}} \to 2 \cdot d_{\mathrm{lat}} \to d_{\mathrm{lat}}$   &            &            & \checkmark \\
        \cmidrule{1-6}
        \multirow{3}{*}{\makecell[l]{Feature map configuration\\ \quad \desc{feature maps per}\\ \quad \desc{convolutional stage in the encoder}}}
             & F-8  & $8  \to 8  \to 8  \to 8$  & \checkmark & \checkmark &            \\
             & F-16 & $8  \to 8  \to 8  \to 16$ &            &            &            \\
             & F-32 & $8  \to 8  \to 16 \to 32$ &            &            & \checkmark \\
        \cmidrule{1-6}
        \multirow{3}{*}{\makecell[l]{Kernel size configuration\\ \quad \desc{kernel size of first encoder and}\\ \quad \desc{last decoder convolutional layer}}}
             & K-3  & $3 \times 3$                     &            & \checkmark &            \\
             & K-5  & $5 \times 5$                     & \checkmark &            &            \\
             & K-7  & $7 \times 7$                     &            &            & \checkmark \\
        \midrule
        CELU parameter $\alpha$ & \multicolumn{2}{l}{range $[1,\,100]$ (log-scaled)}             & $5$       & \multicolumn{2}{c}{\multirow{3}{*}{\makecell{\desc{no effect on the}\\ \desc{number of parameters}}}} \\
        Batch size              & \multicolumn{2}{l}{$\{32, \, 64, \, 128, \, 256, \, 512\}$}            & $256$     & \multicolumn{2}{c}{} \\
        Learning rate           & \multicolumn{2}{l}{range $[10^{-4},\,10^{-2}]$ (log-scaled)}   & $10^{-3}$ & \multicolumn{2}{c}{} \\
        \bottomrule
    \end{tabular}
\end{table*}

\subsubsection{CAE training}
\label{sec:methods:dimension_reduction:cae_training}

The reconstruction error is minimised using Adam (\emph{adaptive moment estimation})~\cite{Kingma2014}, an adaptive variant of \emph{stochastic gradient descent} (SGD). The SGD method implies that the parameter gradients are estimated based on batches rather than the full dataset~\cite{Kingma2014} with a batch size of $T_{\mathrm{Batch}} = 256$ snapshots in this case. The learning rate was set to $10^{-3}$ based on the results of the hyperparameter studies (see~\ref{app:hyperparameter_optimisation}).

\begin{figure}[tb]
    \centering
    \includegraphics{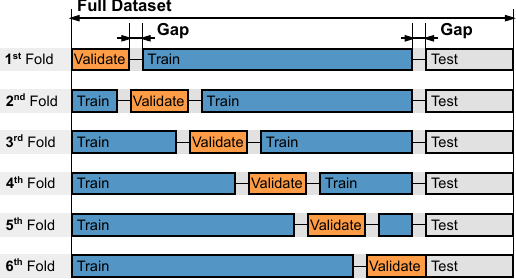} 
    \caption{\textbf{Illustration of the six-fold training-validation-test split.} Temporal gaps are enforced between the validation subset and the training subset to prevent temporal leakage.}
    \label{fig:Validation_split}
\end{figure}

The training is performed using part of the full dataset, the training subset, while being monitored using another part of the reference dataset, the validation subset. The validation subsets are defined following the six-fold cross-validation method as illustrated in Figure~\ref{fig:Validation_split}, whilst specific subset sizes are given in Section~\ref{sec:methods:datasets}. Reconstruction errors $e_{\mathrm{NMSE}}$ are reported for an unseen test subset. To prevent information leakage from the validation subset to the training subset due to temporal correlations between consecutive snapshots, a temporal gap is enforced before and after the training subset, which is deducted from the training subset. This temporal gap is inferred from the largest integral length scale in the POD time series.

To prevent overfitting, the training process is stopped at the point of the lowest achieved reconstruction error on the validation set (i.e., the early stopping method).

\subsection{Non-linearity metrics}
\label{sec:methods:metrics}

This paper proposes two non-linearity metrics: the LTR is applied \emph{a priori} to autoencoder training to estimate how suitable non-linear compression is for a given dataset. The DLE is used \emph{a posteriori} to measure the actual non-linearity captured in the trained autoencoder.

\subsubsection{Local tangent residual (LTR)}
\label{sec:methods:metrics:ltr}

\begin{figure*}[tb]
    \centering
    \includegraphics{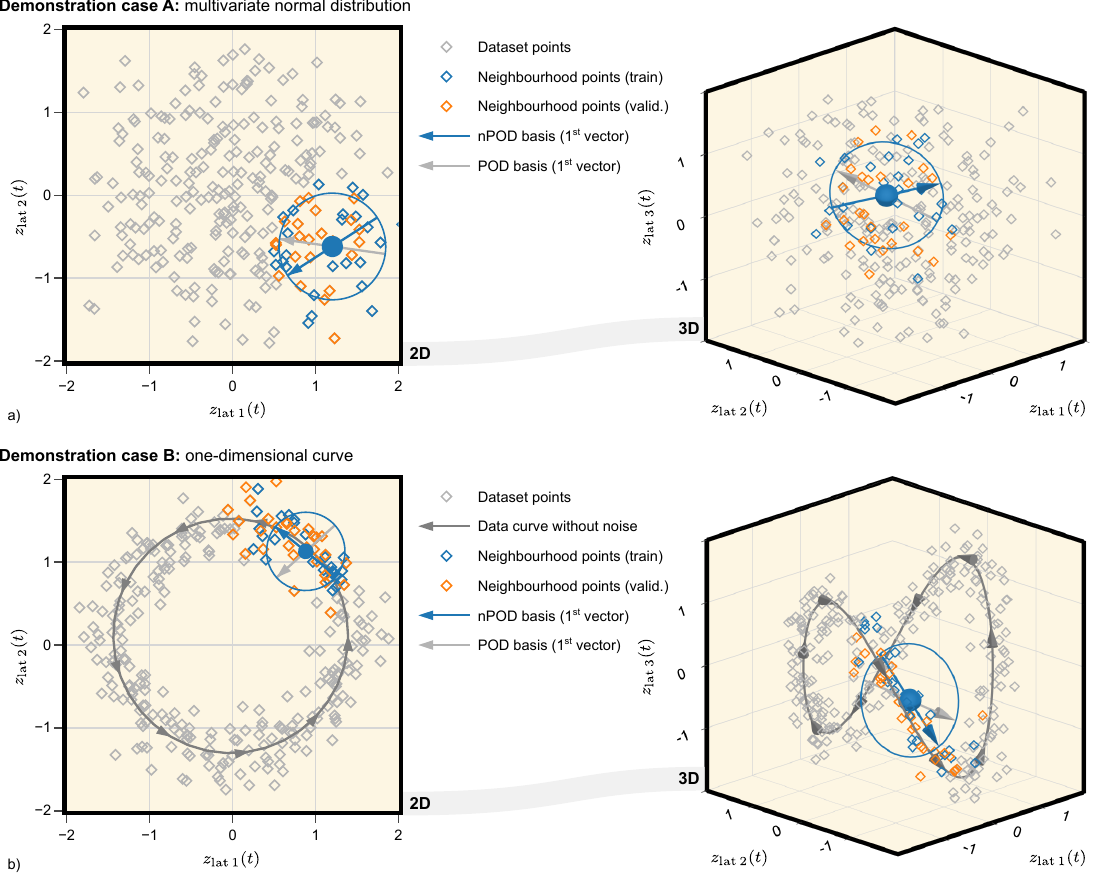}
    \caption{\textbf{Schematic illustrating the LTR metric.} The data points, divided into training and validation points, are highlighted for a single neighbourhood. The nPOD basis is calculated using the training points in the neighbourhood. The POD basis is computed for all training points across the entire dataset. \textbf{(a)} Data points of a multivariate normal distribution. \textbf{(b)} Data points of a one-dimensional curve with added noise.}
    \label{fig:LPOD_Demonstration_2D}
\end{figure*}
 
The concept behind the LTR is to quantify the extent to which data points can be better represented by a \emph{local tangential} coordinate system compared to the \emph{global} orthonormal coordinate system defined by the POD modes. The \emph{local residual} between tangential representation and the global orthonormal coordinate system is then interpreted as the potential for an autoencoder to find a more efficient latent encoding compared to POD.

Before computing the metric, all points in the dataset are split into a \emph{training} subset and a \emph{validation} subset. The method then defines multiple local \emph{neighbourhoods} $\mathcal{N}$ of points by drawing random \emph{anchor} points from the training subset and selecting their respective $k_{\mathrm{train}}$ nearest neighbours in the training subset. The $k_{\mathrm{valid}}$ validation points are assigned to the neighbourhoods afterwards: starting from the known anchor point, the nearest neighbours from the validation set are added until the average distance of the validation points from the anchor point approximately matches the average distance of the training points from the anchor point. The latter average distance is hence interpreted as the radius of the neighbourhood $r_{\mathcal{N}}$.

To compute the LTR, a POD basis is first derived from all training points of the entire dataset. The method then applies the following algorithm to each neighbourhood $\mathcal{N}$:

\begin{enumerate}
    \item A local POD basis is computed on the training points of the neighbourhood, referred to as the \emph{neighbourhood}-POD (nPOD) basis. Before computing this basis, the neighbourhood is centred, i.e., the mean of its training points is subtracted.

    \item The nPOD basis is truncated to $d_{\mathrm{lat}}$ modes, and the local normalised reconstruction error $e_{\mathrm{NMSE}, \, \mathcal{N}}$ is computed on the validation points of the centred neighbourhood. In this context, the nPOD basis acts as a local tangential representation of the data manifold.

    \item The POD basis is likewise truncated to $d_{\mathrm{lat}}$ modes, and its local reconstruction error $e_{\mathrm{NMSE}, \, \mathcal{N}}$ is computed on the same validation points.
\end{enumerate}

Both local reconstruction errors describe how well the respective bases capture the structure (i.e., the variance) of the validation points in the local neighbourhood: on the one hand, a relatively small error for the nPOD basis compared to a large error for the POD basis suggests a strong non-linear \emph{out-of-sample} compression potential. On the other hand, if the POD basis represents the data points just as well as the nPOD basis, then non-linear compression is not advantageous over linear compression.

\paragraph{Examples for LTR computation and interpretation}
 
The LTR concept is illustrated in Figure~\ref{fig:LPOD_Demonstration_2D}. The datasets shown are two- or three-dimensional, with the aim being to compress them to a single dimension, i.e., $d_\mathrm{lat} = 1$. The corresponding nPOD bases (shown in blue) and POD bases (shown in grey) therefore consist of only a single vector which, by definition, points in the direction of the greatest variance in the data.

Such a 2D/3D dataset is obtained, for example, examining the first POD time series $\latentVariableVecTime{t}$ of a turbulent flow process. The LTR method can be applied both in the high-dimensional space of the snapshots and in the latent space, provided that the latent space is generated from a linear mapping (via POD) and is sufficiently large to represent the relevant dynamics; after all, the structure of the data points under investigation must be preserved in the latent space during this dimension reduction.

Case (a), where data points are sampled from a multivariate normal distribution, exemplifies an unstructured dataset. On average, the nPOD and POD bases in the local neighbourhoods represent the same low proportion of the variance here. 

The situation is different for case (b), which consists of data points on a one-dimensional curve with superimposed noise, i.e., a structured dataset. Despite the noise, the nPOD bases do, on average, capture the variance in the data better. If the data-generating curve is interpreted as a curvilinear coordinate system (e.g., a polar coordinate system in the 2D case), then the first coordinate dimension can represent a much larger proportion of the variance. This consideration is relevant in the context of fluid flow, as it is known that, for example, a \textsc{Kármán} vortex street follows such a curve in the POD latent space~\cite{Noack2003}.

The difference in local reconstruction errors, averaged across all local neighbourhoods with the same number of training points $k_{\mathrm{train}}$, remains a function of the mean neighbourhood radius $\overline{r}_{\mathcal{N}}$. We define the LTR as the maximum of this difference over the neighbourhood radius:

\begin{equation}
    \label{eqn:LTR_definition}
    e_{\mathrm{LTR}}(d_{\mathrm{lat}}) = \max_{\overline{r}_{\mathcal{N}}} \, \Delta\overline{e_{\mathrm{NMSE}, \, \mathcal{N}}}(\overline{r}_{\mathcal{N}}, \, d_\mathrm{lat}) \, .
\end{equation}

\begin{figure}[tb]
    \centering
    \includegraphics{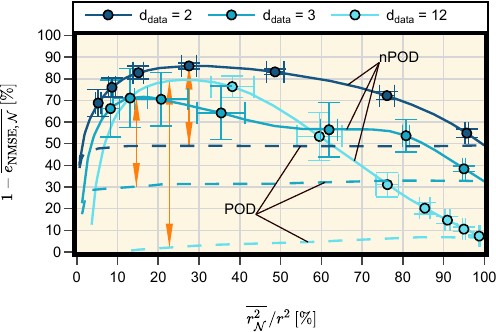}
    \caption{\textbf{Example of how to compute the LTR.} The plot shows the averaged local reconstruction fraction for both the nPOD (solid lines) and the POD basis (dashed lines) against the normalised mean squared radius of the neighbourhoods. Three different benchmark cases are shown with different data dimensions. Error bars denote the standard deviation across the different neighbourhoods.}
    \label{fig:NMSE_LPOD_Benchmark}
\end{figure}

This procedure is illustrated in Figure~\ref{fig:NMSE_LPOD_Benchmark} for three different periodic benchmark datasets. The first two datasets correspond to case (b) discussed above and shown in Figure~\ref{fig:LPOD_Demonstration_2D}, with the same number of data points and the same noise level. The data points for these benchmark cases ($d_{\mathrm{data}} \in \{2, \, 3, \, 8\}$) were generated according to a simple construction rule:

\begin{equation}
    \latentVariableVecTime{t} = \rho \cdot \begin{bmatrix} \cos(t) \\ \sin(t) \\ \cos(2t) \\ \sin(2t) \\ \cos(3t) \\ \sin(3t) \\ \vdots\end{bmatrix} \in \mathbb{R}^{d_{\mathrm{data}}}
\end{equation}

\noindent where the radius $\rho$ was superimposed with \textsc{Gaussian} noise ($\sigma_{\mathrm{noise}} = 0.25$).

Figure~\ref{fig:NMSE_LPOD_Benchmark} shows the averaged local reconstruction fraction plotted against the \emph{normalised mean squared neighbourhood radius}\footnote{Rather than the mean radius, the data is plotted against the mean squared radius of the neighbourhoods. This improves the visual resolution of larger neighbourhoods.}. Here, the normalisation constant is the ``total radius'' of the entire dataset (i.e., the total variance in all dimensions, $r^{2} = \mathrm{Var}_{\mathrm{total}}$). Equivalently, the \emph{normalised mean squared neighbourhood radius} can be interpreted as the ratio of the total variance within a neighbourhood to that of the entire dataset. The reconstruction fraction for all data points can therefore be read on the right-hand side of the plot (i.e., at $\overline{r_{\mathcal{N}}^2} / r^2 = 100 \, \%$), where the neighbourhood contains the entire dataset and thus nPOD and POD coincide at this point.

The nPOD curve can be divided into two regions: (1) for small neighbourhoods, the error is high. This is due to local \emph{overfitting}, as in this region the structure of the training points no longer corresponds to that of the validation points. (2) For large neighbourhoods, the error is also high, as the nPOD bases do not precisely capture the local structure. Between these two regions, the nPOD local reconstruction fraction reaches a maximum, which also lies close to the maximum difference between nPOD and POD. Here, the nPOD basis represents the data points significantly better on average than the POD basis. The curves of the POD basis cannot adapt to the local geometry of the data points and therefore show an almost constant reconstruction result across all neighbourhood sizes. 

\paragraph{Additional considerations and assumptions}

The LTR is defined as a local metric: its local reconstruction errors are computed relative to the total variance within each local neighbourhood. This definition is chosen in order to construct a metric that is free of parameters other than the desired number of latent dimensions. However, this definition also implies that the metric’s significance is limited to measuring the local reconstruction error instead of the reconstruction error for all data points. We nevertheless interpret the LTR as an estimator here to predict a priori whether a capable and well-trained autoencoder (or comparable non-linear dimension reduction methods) will be able to outperform POD on the entire dataset. Hence, a large LTR suggests non-linear compression potential for the autoencoder.

For this interpretation to be meaningful, we need to assume that the autoencoder meets certain requirements. These requirements are derived directly from the tangential modelling: The nPOD basis provides a local approximation of an encoder--decoder pair that is linear and hence \textsc{Lipschitz} continuous\footnote{A mapping $f_{\parameterSetEncoder}$ is \textsc{Lipschitz} continuous with a \textsc{Lipschitz} constant $L$ if there exists a constant $L \geq 0$ such that $\bigl\|f_{\parameterSetEncoder}\bigl(\stateMatrixTime{t_1}\bigr) - f_{\parameterSetEncoder}\bigl(\stateMatrixTime{t_2}\bigr)\bigr\|_2 \leq L \, \bigl\|\stateMatrixTime{t_1} - \stateMatrixTime{t_2}\bigr\|_2$ for all $\stateMatrixTime{t_1}$ and $\stateMatrixTime{t_2}$ (and analogously for $g_{\parameterSetDecoder}$ in the latent space).} with a \textsc{Lipschitz} constant one. Furthermore, this tangential approximation is continuous under small variations of the neighbourhood. The corresponding approximated encoder--decoder pair is therefore not arbitrarily sensitive to small perturbations in the input. For the LTR to provide a meaningful estimate of what a trained autoencoder could achieve, the autoencoder is assumed to accordingly exhibit a continuously differentiable encoder--decoder pair with a bounded \textsc{Lipschitz} constant of comparable order.

The reason for this requirement becomes apparent when considering a space-filling curve (e.g., a \textsc{Peano} curve), which provides a continuous parametrisation of all points in an $n$-dimensional space using a single coordinate. An autoencoder that mapped all points onto such a curve could, in principle, represent the entire dataset using a single latent dimension. However, such a mapping is not \textsc{Lipschitz} continuous and hence does not satisfy the conditions above. Nevertheless, it makes sense not to consider these cases, as such a parametrisation would be sensitive to input perturbations (i.e., a strong overfitting), rendering meaningful interpolation and physical interpretation of the latent space impossible.

In addition, we assume that the nPOD bases can in fact be combined into a single global coordinate system. A low reconstruction error within each individual neighbourhood does not in itself guarantee that the local tangential coordinate systems can be assembled across neighbourhood boundaries into such a global system that satisfies the conditions stated above.

\subsubsection{Decoder linearisation error (DLE)}
\label{sec:methods:metrics:dle}

The DLE quantifies the non-linearity of the decoder after training based on a linearisation of the decoder mapping function $g_{\parameterSetDecoder}: \mathbb{R}^{d_{\mathrm{lat}}} \to \mathbb{R}^{3 \times \numberOfGridPointsIdx{1} \times \numberOfGridPointsIdx{2}}$ at the origin of the latent space $\boldsymbol{0}_{\mathrm{lat}} \in \mathbb{R}^{d_{\mathrm{lat}}}$ using a first-order \textsc{Taylor} approximation, which serves as a local linear surrogate for the decoder:

\begin{equation}
	\mathcal{T}_{1}g_{\parameterSetDecoder}(\latentVariableVec) := g_{\parameterSetDecoder}(\boldsymbol{0}_{\mathrm{lat}}) + g_{\parameterSetDecoder}^{\prime}(\boldsymbol{0}_{\mathrm{lat}})\latentVariableVec \, ,
\end{equation}

\noindent where $g_{\parameterSetDecoder}^{\prime}(\boldsymbol{0}_{\mathrm{lat}})$ denotes the \textsc{Jacobian} evaluated at the latent origin.

Then, the NMSE between the reconstructions generated using the original decoder $g_{\parameterSetDecoder}$ and the surrogate model $\mathcal{T}_{1}g_{\parameterSetDecoder}$, normalised by the mean dataset TKE, provides a measure of the decoder's local non-linearity under variation of the latent variable $\latentVariableVec$:

\begin{equation}
	e_{\mathrm{DLE}}(\latentVariableVec) = \frac{\bigl\|g_{\parameterSetDecoder}(\latentVariableVec) - \mathcal{T}_{1}g_{\parameterSetDecoder}(\latentVariableVec)\bigr\|_{2}^{2}}{\frac{1}{T}\sum_{t \in \mathcal{T}} \bigl\|\stateMatrixTime{t}\bigr\|_{2}^{2}} \, .
\end{equation}

Furthermore, the variation of each single latent component (i.e., $\latentVariableVec = \delta\boldsymbol{\mathrm{e}}_{i} = [0 \, \ldots \, \delta \, \ldots \, 0]$) while keeping the others fixed at zero allows each separate latent dimension to be considered. However, the proposed metric still depends on the latent variable. To eliminate this influence, a weighting by the density of the standard normal distribution over the latent variable is applied. This weighting reflects that the encoder architecture enforces unit variance in each dimension of the latent variable via batch normalisation without trainable parameters. The integral over this weighted term finally provides a scalar non-linearity measure for each latent component:

\begin{align}
	[e_{\mathrm{DLE}}]_{i} &= \int_{-\infty}^{\infty} p(\delta) e_{\mathrm{DLE}}(\delta\boldsymbol{\mathrm{e}}_{i}) \, \mathrm{d}\delta \, ,\\
    p(\delta) &:= \tfrac{1}{\sqrt{2\pi}} \, \exp\bigl({-\tfrac{1}{2}\delta^2}\bigr) \, .
\end{align}

\subsection{Datasets}
\label{sec:methods:datasets}

\begin{figure}[tb]
    \centering
    \includegraphics{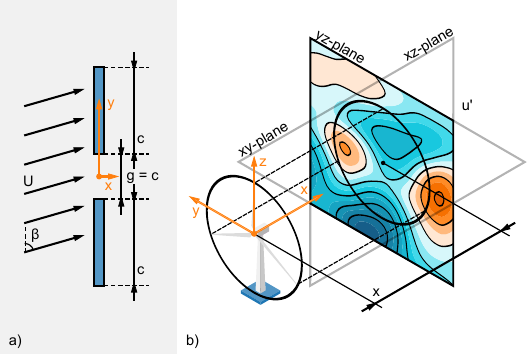}
    \caption{\textbf{Sketches of the setup and coordinate systems for the different flow cases.} \textbf{(a)} Collinear plates setup for the reference datasets. \textbf{(b)} Snapshot planes in the wind turbine wake.}
    \label{fig:Dataset_setup}
\end{figure}

Three datasets were examined in this work: first, two different \emph{reference cases} that consider the viscous flow around two collinear flat plates with chord length $c$ in the low-\textsc{Reynolds} number regime ($\mathrm{Re}_{\mathrm{c}} = \frac{U_{\infty}c}{\nu} = 40$ for the first reference case and $\mathrm{Re}_{\mathrm{c}} = \frac{U_{\infty}c}{\nu} = 100$ for the second reference case)~\cite{SoleraRico2024a}. The third dataset contains snapshots from different planes in the wake of a wind turbine with diameter $D$. This wake flow is characterised by a high \textsc{Reynolds} number ($\mathrm{Re}_{\mathrm{D}} = \frac{U_{\infty}D}{\nu} \approx 10^8$). Figure~\ref{fig:Dataset_setup} shows a diagram of the flow cases in question.

\subsubsection{Reference datasets}
\label{sec:methods:datasets:reference_datasets}

The open-access reference datasets that were used in this work were created and discussed in the ROM context~\cite{Asztalos2023, SoleraRico2024a} and were published via \emph{Zenodo}~\cite{SoleraRico2024}. The following description is based on these references.

The datasets comprise simulation results for the incompressible, two-dimensional, viscous flow over two flat plates of (single-plate) chord length $c$ that are collinear and arranged with an angle $\beta$ to the oncoming flow with a gap of $c$ (illustrated in Figure~\ref{fig:Dataset_setup}). Two distinct flow regimes are distinguished:

\begin{enumerate}
    \item Flow case with perpendicular inflow ($\beta = 90^{\circ}$), a \textsc{Reynolds} number of $\mathrm{Re}_{\mathrm{c}} = 40$ and comprising $1000$ flow field snapshots ($200$ snapshots removed due to transients, $625$ snapshots in training subset minus gaps of $50$ snapshots, $125$ snapshots in validation subset and $50$ snapshots in test subset for one fold).
    \item Flow case with non-perpendicular inflow ($\beta = 80^{\circ}$), a \textsc{Reynolds} number of $\mathrm{Re}_{\mathrm{c}} = 100$ and comprising $150 \, 000$ flow field snapshots ($100 \, 000$ snapshots in training subset minus gaps of $50$ snapshots, $20 \, 000$ snapshots in validation subset and $30 \, 000$ snapshots in test subset for one fold).
\end{enumerate}

The simulations are based on direct numerical simulation with an \emph{immersed-boundary projection method} (IBPM) solved on three nested uniform grids with a total domain size of $96c \times 28c$. The original simulation data is downsampled to a spatial resolution of $300 \times 98$ grid points with a temporal resolution of one convective time $\Delta t = \frac{c}{U_{\infty}} = t_{\mathrm{c}}$ in the first case ($\mathrm{Re}_{\mathrm{c}} = 40$) and $\Delta t = \frac{1}{5} \, t_{\mathrm{c}}$ in the second case ($\mathrm{Re}_{\mathrm{c}} = 100$). To ensure that the reference datasets are consistent with the wind turbine wake case and to avoid the need to adapt the CAE architecture, a subset of the data with a spatial resolution of $48 \times 48$ grid points was used.

\begin{figure}[tb]
    \centering
    \includegraphics{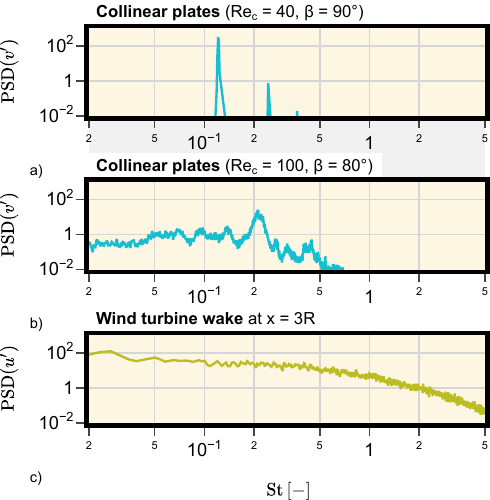}
    \caption{\textbf{PSD plots for the investigated datasets.} \textbf{(a)} Collinear plates case with $\mathrm{Re}_{\mathrm{c}} = 40$ and inflow angle $\beta = 90^{\circ}$ at $\coordX=2c$. \textbf{(b)} Collinear plates case with $\mathrm{Re}_{\mathrm{c}} = 100$ and inflow angle $\beta = 80^{\circ}$ at $\coordX=2c$. \textbf{(c)} Wind turbine wake case (yz-plane subset, $\coordX=3R$, $U_{\mathrm{ref}} = 6 \, \mathrm{m/s}$).}
    \label{fig:dataset_reference_spectrum}
\end{figure}

Figure~\ref{fig:dataset_reference_spectrum}a and Figure~\ref{fig:dataset_reference_spectrum}b present the \emph{power spectral density} (PSD) plots for the fluctuating $v^{\prime}$-component for both cases that were evaluated at a point $\coordX=2c$ downstream of the two collinear plates. The peak of the periodic vortex shedding and its harmonics stand out in the first reference case where the base peak forms at the \textsc{Strouhal} number $\mathrm{St} = 0.12$. This peak is shifted to $\mathrm{St} = 0.21$ for the second reference case with higher \textsc{Reynolds} number.

\subsubsection{Wind turbine wake dataset}
\label{sec:methods:datasets:wind_turbine_wake_dataset}

The wind turbine wake dataset comprises a series of snapshots that were extracted from LES results for a single wind turbine wake flow. To this end, the turbine simulations were carried out in the \emph{EllipSys3D}~\cite{Michelsen1992,Michelsen1994,Soerensen1995} incompressible flow solver, which uses a collocated-grid finite-volume discretisation to solve the \textsc{Navier--Stokes} equations in curvilinear coordinates. In LES, the \textsc{Navier--Stokes} equations are space-filtered, and a subgrid-scale model, the \textsc{Deardorff}~\cite{Deardorff1980} model in this case, resolves the smaller spatial scales. The convective terms are discretised using a third-order QUICK scheme, while the time steps are computed using a second-order implicit method. To solve the pressure correction equation, an improved variant of the SIMPLEC algorithm~\cite{Shen2003} is used, and pressure decoupling is prevented via \textsc{Rhie--Chow}~\cite{Rhie1983} interpolation.

The simulation subject of the wind turbine wake dataset is the IEA $15\, \mathrm{MW}$ reference turbine model~\cite{Gaertner2020} with a radius of $R = 120 \, \mathrm{m}$ and a hub height of $150 \, \mathrm{m}$ that is implemented using the aero-servo-elastic solver HAWC2~\cite{Larsen2025}. The turbine is modelled in EllipSys3D using the \emph{actuator disc method} (ADM)~\cite{Mikkelsen2004}, which transfers aero-elastic forces computed in HAWC2 to the flow field~\cite{Hodgson2021} through the open-source coupling framework DYNAMIKS~\cite{Pedersen2026}.

The simulation was carried out in two stages, with the first stage being the precursor simulation that generates \emph{atmospheric boundary layer} (ABL) flow data without the turbine to provide an atmospheric inflow realisation. This precursor flow data is applied as an inlet condition to the latter successor simulation that includes the turbine. To model the ABL in the precursor simulation, a \emph{pressure-driven boundary layer} (PDBL) model is applied and then simulated for $84 \, \mathrm{h}$ in total on a grid with $512 \times 512 \times 192 \approx 50 \times 10^{6}$ grid points (streamwise, lateral and vertical) and cells of $20 \, \mathrm{m} \times 20 \, \mathrm{m} \times 10 \, \mathrm{m}$. The last $24 \, \mathrm{h}$ are extracted to provide statistically stationary inflow to the successor simulation.

The successor simulation is carried out on a grid with $896 \times 192 \times 128 \approx 22 \times 10^{6}$ grid points (streamwise, lateral and vertical) with a resolution of $20$ cells per turbine radius $R$, corresponding to $44.8 \, R \times 9.6 \, R \times 6.4 \, R$. This setup achieves a high resolution in the ADM context, with an expected error of less than $1 \, \mathrm{\%}$ in relation to the thrust coefficient $C_{\mathrm{T}}$~\cite{Hodgson2021}. The simulation was run for $24 \, \mathrm{h}$\footnote{Transient temperature effects, which are present in the real atmosphere on this time scale, are neglected to generate sufficient data for a statistically stationary flow.}.

\begin{figure}[tb]
    \centering
    \includegraphics{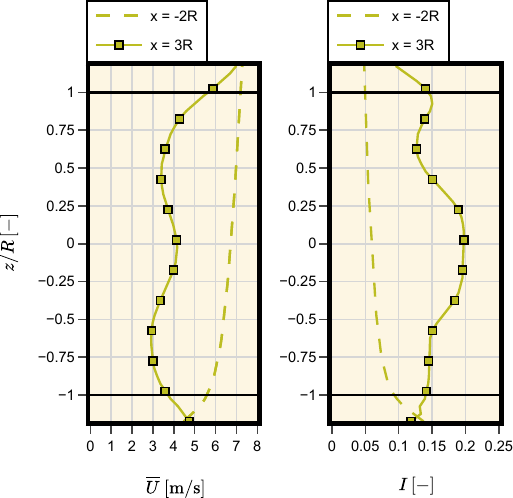}
    \caption{\textbf{Profiles of the mean velocity and the turbulence intensity upstream and downstream of the wind turbine.} The profiles are extracted along vertical lines at the positions $\coordX=-2R$ and $\coordX=3R$. The rotor disc is sketched using black lines.}
    \label{fig:dataset_reference_plots}
\end{figure}

The resulting mean velocity profiles are shown in Figure~\ref{fig:dataset_reference_plots} along with the turbulence intensities. The data were extracted from the successor simulation along one vertical line upstream ($\coordX=-2R$, $\coordY=0$) and downstream ($\coordX=3R$, $\coordY=0$) of the wind turbine. The mean velocity profiles exhibit the characteristic logarithmic wind profile upstream of the turbine, with a wake deficit downstream of the turbine. The turbulence intensity is highest in the region of the wake centre, with two further local maxima in the region of the shear layer from the rotor disc. A representative PSD for this flow case is given in Figure~\ref{fig:dataset_reference_spectrum}c at the same downstream position ($\coordX=3R$, $\coordY=0$, $\coordZ=0$) based on the \textsc{Strouhal} number with a reference wind speed of $U_{\mathrm{ref}} = 6 \, \mathrm{m/s}$. The PSD shows a significantly broader temporal distribution of TKE compared to the two reference flow cases and with no discernible shedding frequency, although meandering~\cite{Yang2019} will occur further downstream.

\begin{figure}[tb]
    \centering
    \includegraphics{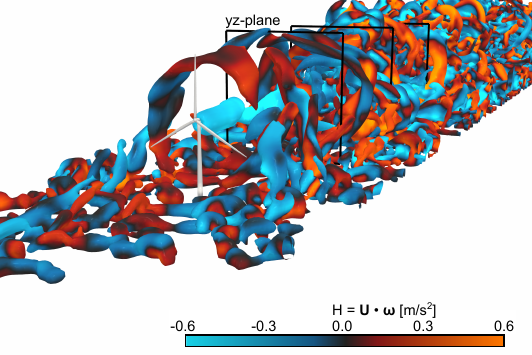}
    \caption{\textbf{Coherent vortex structures in the turbulent wind turbine wake.} The visualisation uses an instantaneous \emph{Q-criterion} isosurface coloured with the helicity to highlight rotational structures. Three yz-planes are sketched at the downstream positions $\coordX \in \{3R, \, 6R, \, 9R\}$.}
    \label{fig:q_criterion}
\end{figure}

In total, a series of $T = 2 \, \mathrm{Hz} \cdot 23 \, \mathrm{h} = 165\,600$ snapshots from the fluctuating flow field $\stateMatrix$ is extracted from the LES results with a temporal resolution of $2 \, \mathrm{Hz}$. These are split into $108 \, 000$ snapshots for the training subset minus temporal gaps of $300$ snapshots (refer to Section~\ref{sec:methods:dimension_reduction:cae_training}), $21 \, 600$ snapshots for the validation subset and $36 \, 000$ snapshots for the test subset for one fold (refer to Figure~\ref{fig:Validation_split}). The first $1 \, \mathrm{h}$ is excluded beforehand to remove possible transients. The snapshots are extracted in three different planes within the wake of the wind turbine: one plane perpendicular to the inflow direction downstream of the wind turbine (yz-plane) and the other two planes containing the flow direction vector (xy-plane and xz-plane). Considering the flow in different planes is intended to enable an investigation of dimension reduction mechanisms from different perspectives: encoding the translational movement of advected turbulent structures in the streamwise plane (xy and xz) versus encoding the changes to the cross-section of such structures in the cross-streamwise (yz) plane. In these planes, the fluctuating velocity components $\boldsymbol{U}^{\prime (t)}$ are extracted at the cell centres within a square spatial domain of $48 \times 48$ grid points at 12 downstream positions in the turbine wake, where $\coordX$ denotes the distance to the centre of these square domains from the turbine on the x-axis. This setup is illustrated in Figure~\ref{fig:Dataset_setup}. Figure~\ref{fig:q_criterion} furthermore provides a visual representation of the wind turbine wake based on the instantaneous Q-criterion isosurface coloured by helicity\footnote{The helicity $H = \mathbf{U} \cdot (\nabla \times \mathbf{U})$ measures the local alignment of velocity and vorticity and thus quantifies the corkscrew motion of the flow.}. The turbine significantly impacts the flow by introducing a circular shear layer and a strong central hub vortex. The shear layer and the hub vortex eventually break down and increase the overall turbulence levels~\cite{Soerensen2015}.

\section{Results}
\label{sec:results}

\subsection{POD modes and eigenvalue distribution}
\label{sec:results:modes_and_eigenvalues}

\begin{figure*}[tb]
    \centering
    \includegraphics{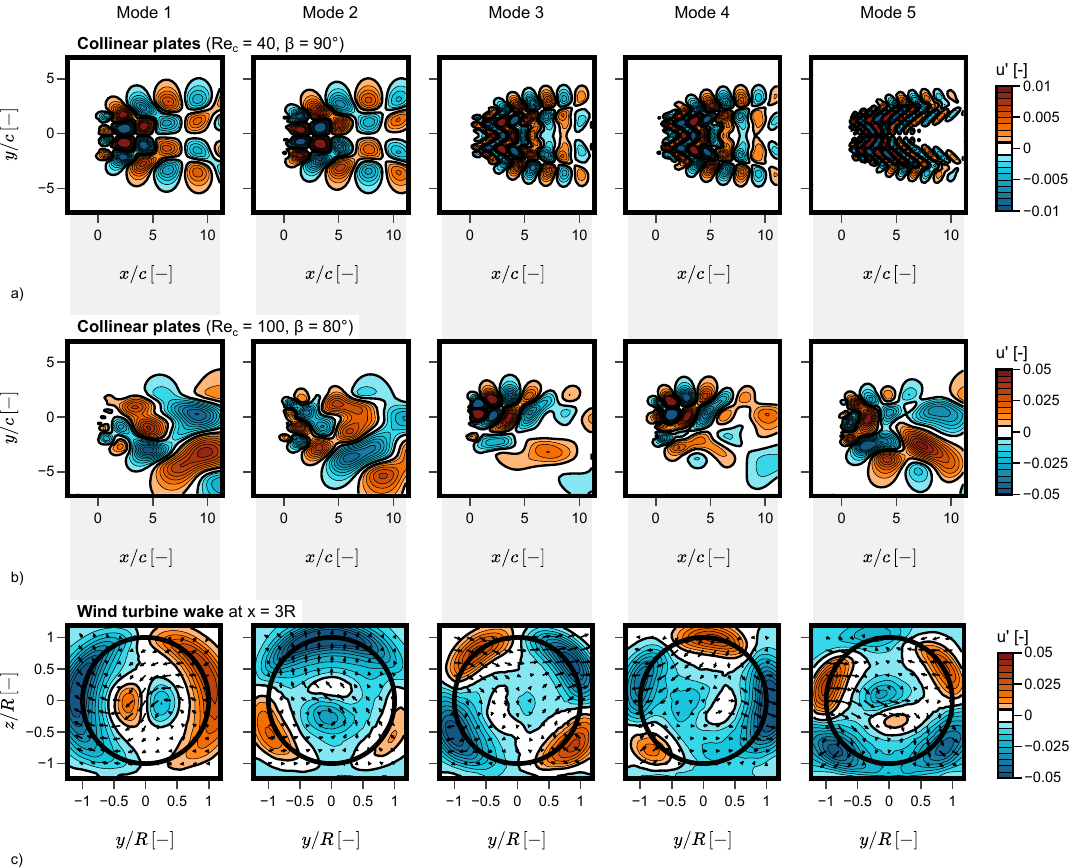}
    \caption{\textbf{Leading POD modes represented as contour plots of the streamwise component.} \textbf{(a)} POD modes for the collinear plates case with $\mathrm{Re}_{\mathrm{c}} = 40$ and inflow angle $\beta = 90^{\circ}$. \textbf{(b)} POD modes for the collinear plates case with $\mathrm{Re}_{\mathrm{c}} = 100$ and inflow angle $\beta = 80^{\circ}$. \textbf{(c)} POD modes for the turbine wake (yz-plane subset, $\coordX = 3R$) with a superimposed quiver plot for the cross-streamwise components.}
    \label{fig:contour_modes}
\end{figure*}

Figure~\ref{fig:contour_modes} shows the leading POD modes for the collinear plates cases and for the wind turbine wake case at the downstream position $\coordX=3R$ for the yz-plane. For all the following experiments, the POD bases were computed for each downstream position and across the three planes individually, ensuring that the POD bases are optimal for each case.

\begin{figure}[tb]
    \centering
    \includegraphics{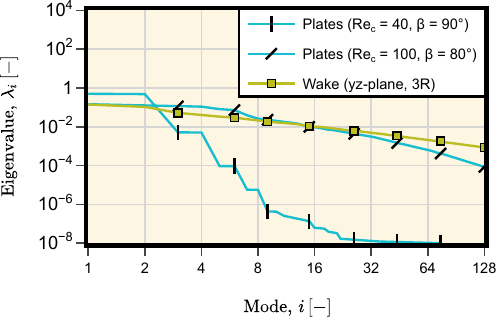}
    \caption{\textbf{Distribution of eigenvalues.} Eigenvalues, normalised to their total sum, for the leading POD modes for the wind turbine wake case ($\coordX=3R$, yz-plane) and the collinear plates cases.}
    \label{fig:eigenvalue_comparison}
\end{figure}

The POD eigenvalue distribution corresponding to these POD modes is shown in Figure~\ref{fig:eigenvalue_comparison}, with the eigenvalues normalised to their total sum. In accordance with the POD definition, the eigenvalues are listed in descending order of magnitude.

In the first collinear plates case, a rapid decay of the eigenvalue distribution is observed. The two dominant leading eigenvalues indicate that only two coherent patterns represent the predominant dynamics. However, the second collinear plates case and the wind turbine wake case exhibit a more broadband distribution of TKE across multiple POD modes and thus spatial scales; there seem to be more coherent patterns of comparable energetic significance. Hence, more POD modes are needed to represent the more complex dynamics for these cases.

Note that the eigenvalues in the first collinear plates case are organised in pairs of two. That is known to be a characteristic POD phenomenon for flows that feature convected coherent structures (e.g., shed vortices). Because POD modes are time-independent, such a dynamic cannot be represented in a single mode. Rather, two modes that are phase-shifted in both space and time need to be superposed to form a travelling wave. That is precisely the case here, with modes 1 and 2 exhibiting this spatial displacement in the flow direction (i.e., the x-direction). The higher modes for the first collinear plates case represent periodic spatial structures that have a wavenumber which is an integer multiple of the wavenumber that is governing the first two modes (see Figure~\ref{fig:contour_modes}a). These higher modes can be associated with the second- and higher-order harmonics that are observed in the PSD for the first collinear plates case (refer to Figure~\ref{fig:dataset_reference_spectrum}a). Hence, the base frequencies along with the higher-order harmonics appear in POD as a series of orthogonal mode pairs, one pair per relevant harmonic.

\subsection{TKE reconstruction}
\label{sec:results:tke_reconstruction}

\begin{figure*}[tb]
    \centering
    \includegraphics{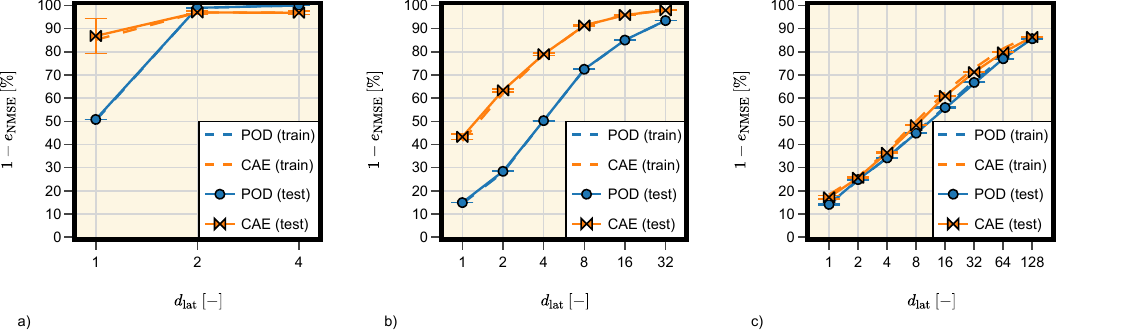}
    \caption{\textbf{TKE reconstruction fraction for different numbers of latent dimensions.} Comparison between CAE and POD for both the training (dashed lines) and the test dataset (solid lines). Error bars denote the standard deviation across the cross-validation folds. \textbf{(a)} Collinear plates case with $\mathrm{Re}_{\mathrm{c}} = 40$ and inflow angle $\beta = 90^{\circ}$. \textbf{(b)} Collinear plates case with $\mathrm{Re}_{\mathrm{c}} = 100$ and inflow angle $\beta = 80^{\circ}$. \textbf{(c)} Wind turbine wake case (yz-plane subset, $\coordX=3R$).}
    \label{fig:NMSE_dlat}
\end{figure*}

Figure~\ref{fig:NMSE_dlat} shows the TKE reconstruction fraction $1 - e_{\mathrm{NMSE}}$, i.e., the fraction of reconstructed TKE in the output relative to the input, for the various datasets and across multiple numbers of latent dimensions (or numbers of POD modes) for both POD and CAE, determined for both the training (dashed line) and the test dataset (solid line).

In the first collinear plates case, the CAE reconstructs around $85 \, \%$ of the flow's TKE with a single latent dimension, whereas POD reconstructs around $50 \, \%$ in a single POD mode. With $d_{\mathrm{lat}} = 2$, both models reconstruct over $90 \, \%$, but POD captures more TKE in comparison. The situation is different in the second collinear plates case, where the CAE requires at least 8 latent dimensions and POD at least 32 to represent $90 \, \%$ of the input TKE. The CAE outperforms POD in the second collinear plates case for all examined numbers of latent dimensions with a large margin as also reported in \cite{SoleraRico2024a}.

The last case examined, the wind turbine wake case (Figure~\ref{fig:NMSE_dlat}c), shows a different trend: the CAE outperforms POD only to a small extent on the test dataset in terms of the reconstruction error, with a maximum difference being achieved at 16 modes. Despite the marginal difference in TKE reconstruction between the two models in this case, the shallow slope of the curves implies that $60 \, \%$, for instance, is achieved with as few as 16 CAE latent dimensions, whereas 22 POD modes are required to attain an equivalent TKE reconstruction fraction. A TKE reconstruction with over $90 \, \%$ of the input test snapshots was not achieved in the range of latent dimensions investigated here.

The observed small differences in reconstruction performance between the training subset and the test subset for both methods have different causes. On the one hand, POD is not prone to overfitting because the linear encoding acts as regularisation per definition. The autoencoder, however, is generally considered to be less robust to overfitting. To mitigate this effect, the hyperparameter studies were carried out in a search space that contained architectures with a wide range of numbers of trainable parameters. Based on the performance on the validation subset, a lightweight model was chosen which, despite its small number of trainable parameters, demonstrates both the best generalisation and low overfitting (refer to~\ref{app:hyperparameter_optimisation}). Consequently, the TKE reconstruction fractions for the training and test subsets are also close to one another for the autoencoder in all three cases.

The definition of POD furthermore enables the eigenvalue distribution to be associated with the observed reconstruction performance: the TKE reconstruction fraction for POD corresponds to the cumulative eigenvalue distribution. For example, the two dominant modes in the first collinear plates dataset lead to a sharp increase in the TKE reconstruction fraction in the first two latent dimensions (Figure~\ref{fig:NMSE_dlat}a); the vast proportion of the TKE in this flow can be reconstructed using only two POD modes. The situation is different in the second collinear plates case (Figure~\ref{fig:NMSE_dlat}b), where the additional angle of attack breaks the flow symmetry and a higher \textsc{Reynolds} number is also enforced. As a result, the flow is more turbulent and also more intermittent, and, consequently, the flow field encoding requires a higher-dimensional latent space to achieve comparable TKE reconstruction. This is reflected in the shallower decay of the eigenvalue distribution relative to the first collinear plates case. The same line of reasoning can also be applied to the wind turbine wake case, which exhibits the broadest spectral TKE distribution (refer to Figure~\ref{fig:dataset_reference_spectrum}c). 

For the autoencoder, the improved reconstruction performance for the collinear plates cases can be linked to the more efficient representation of higher-order harmonics. These appear in POD as separate mode pairs, as observed in the eigenvalue distribution for the first collinear plates case. A non-linear decoder, however, is not restricted to representing each harmonic in separate mode pairs: higher-order harmonics are deterministic but non-linear functions of the fundamental phase (e.g., $\cos(2\theta) = 2\cos^2(\theta)-1$). Therefore, a decoder can reconstruct all relevant harmonics from a single two-dimensional latent representation of the phase.

\begin{figure*}[tb]
    \centering
    \includegraphics{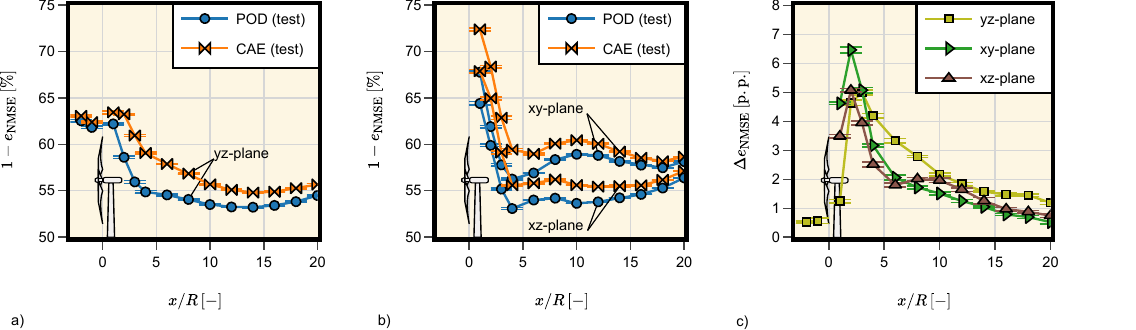}
    \caption{\textbf{TKE reconstruction fraction for the wind turbine wake case at different downstream positions.} Comparison between CAE and POD for a fixed number of latent dimensions $d_{\mathrm{lat}} = 16$ on the test subset. Error bars denote the standard deviation across the cross-validation folds. \textbf{(a)} yz-plane subset. \textbf{(b)} xy-plane and xz-plane subsets. \textbf{(c)} Difference between CAE and POD in percentage points for all subsets.}
    \label{fig:NMSE_downstream}
\end{figure*}

The TKE reconstruction fraction for the wind turbine wake case for different downstream positions and across the various planes (yz-plane, xy-plane and xz-plane) is shown in Figures~\ref{fig:NMSE_downstream}a and~\ref{fig:NMSE_downstream}b for the test dataset and a fixed number of latent dimensions $d_{\mathrm{lat}} = 16$. Figure~\ref{fig:NMSE_downstream}c shows the difference between the CAE and POD TKE reconstruction fractions on the test dataset in percentage points. In general, a trend of decreasing reconstruction performance for both models is observed with increasing distance from the turbine in all planes. Both TKE reconstruction curves have their maximum right behind the turbine ($\coordX=1R$). For the yz-planes, a global minimum is observed at $\coordX=14R$, whereas the streamwise planes feature a local maximum at $\coordX=10R$ for the xy-plane and at $\coordX=8R$ for the xz-plane. The difference curve between CAE and POD shows a pronounced maximum just behind the turbine at $\coordX=3R$ for the yz-plane and $\coordX=2R$ for the other two planes. At this maximum, the CAE has a larger TKE reconstruction fraction with improvements between $4.5 \, \mathrm{p.p.}$ and $6.5 \, \mathrm{p.p.}$ compared to POD.

The observations for POD in the yz-plane subset have been reported and discussed before by \textsc{Sørensen} et al.~\cite{Soerensen2015}, where the minimum in the TKE reconstruction curve is associated with the end of the near-wake region; after the breakdown of the coherent turbulent structures, namely the concentrated tip and root vortices, more POD modes are required to represent the same amount of TKE. Further downstream, the flow recovers from the influence of the turbine and the TKE content of the POD modes gradually returns to its initial level.

The non-linear compression of the autoencoder is very limited (i.e., comparable to POD) in the freestream inflow and the far-wake ($\coordX>14R$). The autoencoder does outperform POD in terms of TKE for $\coordX<14R$, albeit only to a small extent. The TKE reconstruction curves show similar trends for both POD and CAE across all plane subsets, with similar maxima. Hence, it cannot be substantiated that the non-linear encoding of streamwise advection offers significant advantages over the cross-streamwise approach.

Most notable in the overarching context is that the reconstruction performance of the CAE constitutes a significant improvement over POD in the reference cases (except in the aforementioned situation in the first collinear plates case) with $d_{\mathrm{lat}, \, \mathrm{CAE}} < d_{\mathrm{lat}, \, \mathrm{POD}}$ for comparable reconstruction errors, although the outperformance appears to be marginal in the wind turbine wake case. This substantial performance gap is remarkable, not only because the same CAE architecture (with identical spatial input and output resolution) and the same optimisation algorithm were used in all cases but also because the specific CAE architecture and parameters of the training process were specifically tuned to the wind turbine wake case in two hyperparameter studies with considerable effort. Although the proposed autoencoder-based ROM was developed for the wind turbine wake case, significantly better performance with respect to the baseline POD is observed in the reference cases.

\subsection{LTR metric}
\label{sec:results:ltr}

To put these results into perspective, the LTR was computed using the average across a minimum of $n_{a} = 32$ anchor points and considering neighbourhoods of at least $k_{\mathrm{valid}} = 10$ validation points. These neighbourhoods were not identified among the high-dimensional snapshots but among the latent POD coordinates (i.e., the modal amplitudes) with 128 dimensions, covering at least 80\% of the actual TKE.

\begin{figure*}[tb]
    \centering
    \includegraphics{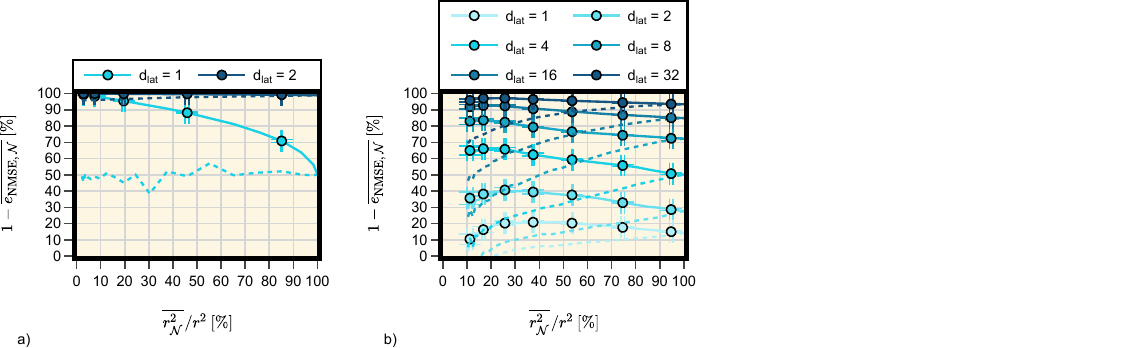}
    \caption{\textbf{LTR plots for the collinear plates cases.} Averaged local reconstruction fraction for the nPOD (solid lines) and POD basis (dashed lines) for different numbers of latent dimensions and neighbourhood radii. Error bars denote the standard deviation across the six cross-validation folds. \textbf{(a)} Collinear plates case with $\mathrm{Re}_{\mathrm{c}} = 40$ and inflow angle $\beta = 90^{\circ}$. \textbf{(b)} Collinear plates case with $\mathrm{Re}_{\mathrm{c}} = 100$ and inflow angle $\beta = 80^{\circ}$.}
    \label{fig:LPOD_reference_cases}
\end{figure*}

The LTR computation is shown in Figure~\ref{fig:LPOD_reference_cases} for the collinear plates cases. These plots present the averaged local reconstruction fraction ($1 - \overline{e_{\mathrm{NMSE}, \, \mathcal{N}}}$) for the nPOD basis (solid line) and the POD basis (dashed line) for different neighbourhood radii $r_{\mathcal{N}}$ and for different numbers of latent dimensions $d_{\mathrm{lat}}$. To standardise the presentation, the curves are plotted against the \emph{normalised mean squared neighbourhood radius} $\overline{r_{\mathcal{N}}^{2}} / r^{2}$, where the normalisation constant is the ``total radius'' of the entire dataset (i.e., the total variance in all dimensions, $r^{2} = \mathrm{Var}_{\mathrm{total}}$).

The first collinear plates case shows a clear maximum in the averaged local reconstruction fraction for the nPOD basis for a small neighbourhood radius (at $\overline{r_{\mathcal{N}}^{2}} / r^{2} \approx 5 \, \%$) of almost $100 \, \%$, which decreases monotonically as the neighbourhood size $r_{\mathcal{N}}$ increases. In the second collinear plates case, the situation is more nuanced, with the local reconstruction fraction of the nPOD bases reaching a local maximum for a larger normalised neighbourhood radius (at $\overline{r_{\mathcal{N}}^{2}} / r^{2} \approx 25 \, \%$). These local maxima are larger than the respective reconstruction fractions for all data points (i.e., for $\overline{r_{\mathcal{N}}^{2}} / r^{2} = 100 \, \%$), indicating a lower \emph{intrinsic dimensionality} and thus the presence of a lower-dimensional structure. This argument is based on the observation that the nPOD basis represents the data points in the neighbourhood more compactly on average (i.e., with a lower $e_{\mathrm{NMSE}}$) than the POD basis can represent all data points.

The POD basis reconstructs a somewhat constant $50 \, \%$ of the input TKE in the first collinear plates case, while in the second collinear plates case the performance increases with larger neighbourhood radii. This latter observation is a probable consequence of the increasing \emph{signal-to-noise ratio} with larger neighbourhoods.

Comparing the nPOD and POD bases in the local neighbourhoods, at a comparable \emph{signal-to-noise ratio}, the nPOD basis yields a more compact representation of the data points within the neighbourhood relative to the POD basis and thus a large LTR. This finding is consistent with the significantly improved reconstruction performance of the CAE with respect to POD for the reference datasets.

\begin{figure*}[tb]
    \centering
    \includegraphics{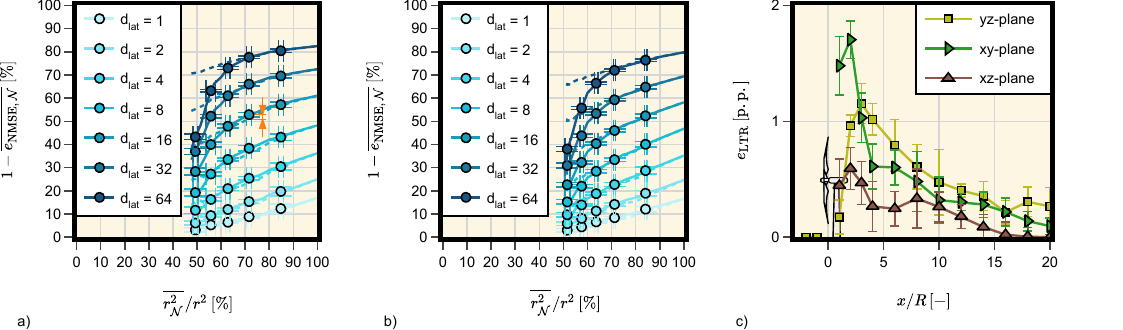}
    \caption{\textbf{LTR plots for the wind turbine wake case.} Averaged local reconstruction fraction for the nPOD (solid lines) and POD basis (dashed lines) for different numbers of latent dimensions and neighbourhood radii. Error bars denote the standard deviation across the six cross-validation folds. \textbf{(a)} xy-plane subset, $\coordX=2R$. The highlighted distance defines the scalar LTR for this dataset for $d_{\mathrm{lat}} = 16$. \textbf{(b)} xy-plane subset, $\coordX=20R$. \textbf{(c)} LTR at different downstream positions in percentage points across all planes for $d_{\mathrm{lat}} = 16$.}
    \label{fig:LPOD_downstream}
\end{figure*}

The LTR computation for the turbine wake case is shown in Figure~\ref{fig:LPOD_downstream}, with Figure~\ref{fig:LPOD_downstream}a showing the plot for the xy-plane $\coordX=2R$ behind the turbine and Figure~\ref{fig:LPOD_downstream}b showing the plot for the xy-plane $\coordX=20R$ behind the turbine. These two positions correspond to the maximum and minimum of the difference curves, as presented in Figure~\ref{fig:NMSE_downstream}c. The LTR, defined as the maximum difference under variation of the neighbourhood size $k$, is shown in Figure~\ref{fig:LPOD_downstream}c for $d_{\mathrm{lat}} = 16$ POD modes.

In the wind turbine wake case there is no clear indicator for a lower-dimensional manifold with the nPOD basis performing worse in terms of the $e_{\mathrm{NMSE}}$ compared to the POD basis that represents all data points. This result is presumed to be attributable to the lower \emph{signal-to-noise ratio} in the local neighbourhoods, with the noise masking a potential manifold structure\footnote{The dataset is examined through a ``magnifying glass'' via the local neighbourhoods: the structure being sought might be hidden in the amplified noise, as the signal-to-noise ratio decreases as the neighbourhood size becomes smaller.}. Despite this, it can be observed that nPOD still outperforms POD with a small margin and thus a small LTR behind the turbine (e.g., at $\coordX=3R$, yz-plane). Within a greater distance to the turbine (at $\coordX=20R$), this potential vanishes with zero LTR.

The LTR thus replicates the trends observed in Figure~\ref{fig:NMSE_downstream}c regarding the difference in reconstruction performance between CAE and POD. Specifically, the maximum of the three difference curves (at $\coordX=3R$ for the yz-planes, and at $\coordX=2R$ for the xy- and xz-planes) is predicted at the correct downstream position, the decay following the maximum is captured accurately, and the xy-plane is correctly identified as exhibiting the greatest compression potential. Furthermore, the LTR curve for the xy-plane reproduces the local minimum at $\coordX=6R$, although the overall predicted magnitude is systematically underpredicted. Here, the underprediction is a consequence of interpreting local reconstruction errors as a measure of the model's performance across the entire dataset.

\subsection{DLE metric}
\label{sec:results:dle}

The LTR quantifies the potential for a non-linear compression from the data alone, prior to model training. If the autoencoder exploits this potential, its decoder must exhibit corresponding non-linear mapping wherever the LTR is large. To assess whether this is the case, the DLE is proposed in this work as a complementary a posteriori metric: it quantifies the deviation between the decoder mapping and its linearised surrogate, thereby measuring the non-linearity realised in the trained decoder.

\begin{figure*}[tb]
    \centering
    \includegraphics{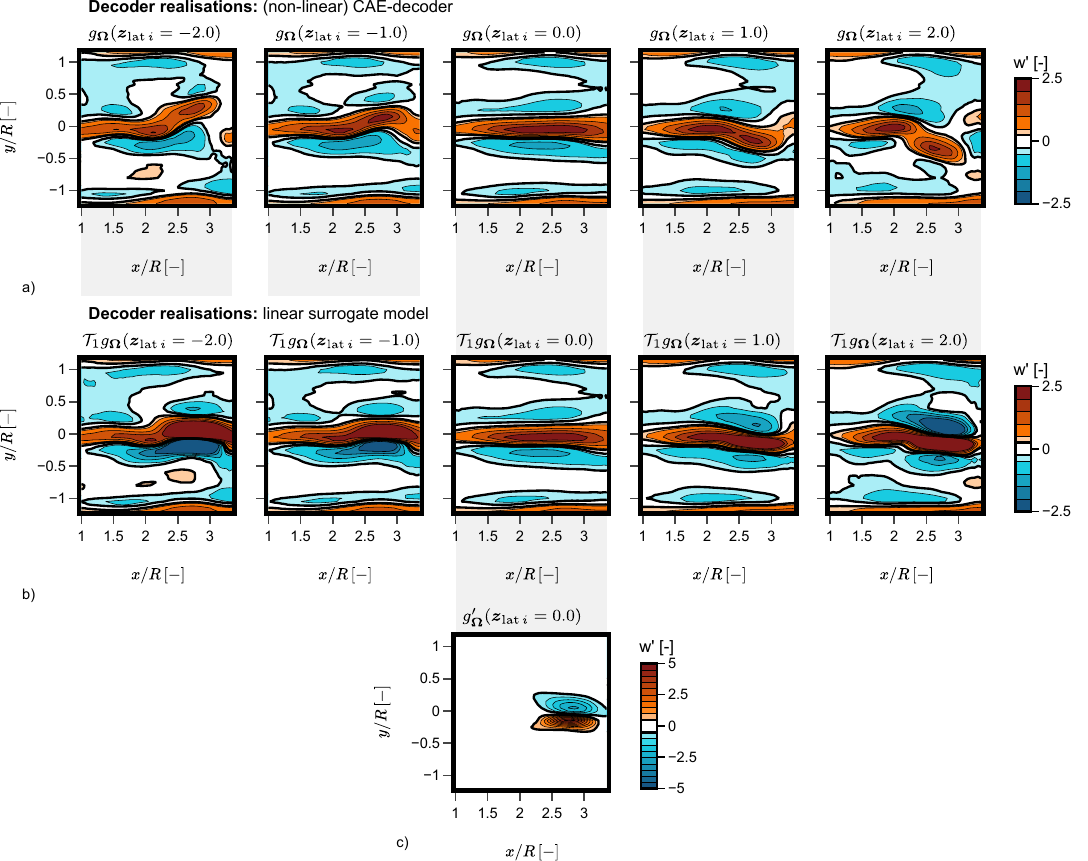}
    \caption{\textbf{Concept of the DLE metric.} Depiction of outputs of the trained decoder and the linear surrogate under systematic variation of the $i$-th dimension of the latent variable. The NMSE between both outputs forms the basis of the DLE. The decoder was trained on the wind turbine wake case at $\coordX=3R$ (xy-plane) with $d_{\mathrm{lat}} = 16$ latent dimensions. \textbf{(a)} Realisations from the trained decoder. \textbf{(b)} Realisations from the linear surrogate model. \textbf{(c)} Gradient evaluated at the origin of the latent space.}
    \label{fig:NLM_IEA15MW_Realisations}
\end{figure*}

To illustrate the DLE concept, Figure~\ref{fig:NLM_IEA15MW_Realisations} presents realisations of a trained decoder and its linear surrogate model under variation of a single latent variable  for the wind turbine wake case (xy-plane at downstream position $\coordX=3R$) along with the gradient evaluated at the origin of the latent space. The two models differ in how the hub vortex motion is encoded: whereas the linear surrogate produces a characteristic dipole, the non-linear decoder is capable of encoding the translational movement that is displacing the region of positive fluctuation velocity. This specific hub vortex motion can also be observed in Figure~\ref{fig:q_criterion} where the hub vortex forms a coherent structure with negative helicity behind the wind turbine nacelle.

This encoding is only possible through a non-linear mapping: since linear mappings are limited to explaining the dynamics through static patterns that are multiplied with the variable, they cannot represent translation, which would need multiple patterns. In other words, flow features that are advected follow curved paths in the high-dimensional snapshot space. The dipole that is produced in the linear surrogate is precisely the first-order (small-amplitude) approximation of this translation; it constitutes a good representation around the origin but is incapable of reproducing the actual motion for larger amplitudes.

\begin{figure}[tb]
    \centering
    \includegraphics{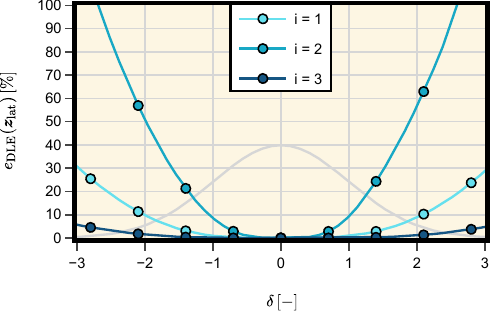}
    \caption{\textbf{DLE metric for different latent dimensions.} The plot presents the NMSE between the reconstructions of the trained decoder and the linear surrogate under variation of the $i$-th latent dimension. The standard normal density used for weighting is shown in the background. The decoder was trained on the wind turbine wake case (xy-plane, $\coordX=3R$) with $d_{\mathrm{lat}} = 16$ latent dimensions.}
    \label{fig:NLM_Channels}
\end{figure}

Figure~\ref{fig:NLM_Channels} further illustrates the computation of the DLE by showing the difference in terms of $e_{\mathrm{NMSE}}$ between the decoder and the linear surrogate across the first three latent dimensions. The normal distribution density function, as shown in the background, is used for pointwise weighting prior to the integration that forms the scalar DLE.

\begin{figure}[tb]
    \centering
    \includegraphics{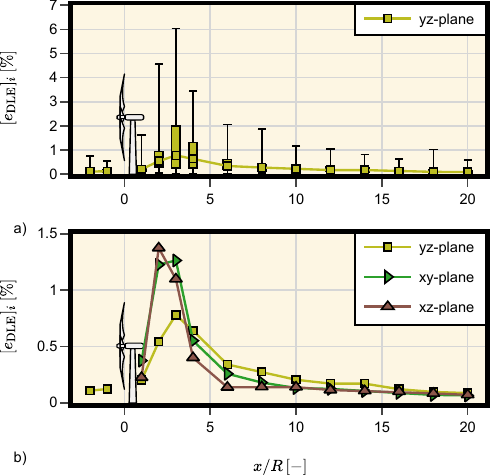}
    \caption{\textbf{DLE metric for different downstream positions.} Representation of DLE values for an autoencoder with $d_{\mathrm{lat}} = 16$ latent dimensions trained on the wind turbine wake dataset. Each boxplot and corresponding median is computed from combined DLE values for six decoders (from six-fold cross-validation). \textbf{(a)} Boxplot representing the lower and upper quartiles of the DLE values for the yz-plane subset at all downstream positions. \textbf{(b)} Median for all subsets and downstream positions.}
    \label{fig:NLM_Downstream}
\end{figure}

The distribution of the DLE values for the wind turbine wake case is represented in Figure~\ref{fig:NLM_Downstream} for $d_{\mathrm{lat}} = 16$ latent dimensions at various downstream positions. For each trained CAE, a single DLE value per latent dimension is computed. Furthermore, because cross-validation involved training six different decoders per flow case, this results in a total of $n_{\mathrm{DLE}} = 6\,d_{\mathrm{lat}}$ scalar DLE values per position and plane. The median of this distribution reaches a maximum at a short distance downstream of the turbine and decreases as the distance from the turbine increases. At the same time, the difference between the lower and the upper quartile of the DLE distribution is also largest behind the turbine. This agreement between the LTR and DLE trends with regard to the position of the maximum suggests that the CAE does indeed exploit the non-linear potential identified a priori using the LTR metric.

\begin{figure}[tb]
    \centering
    \includegraphics{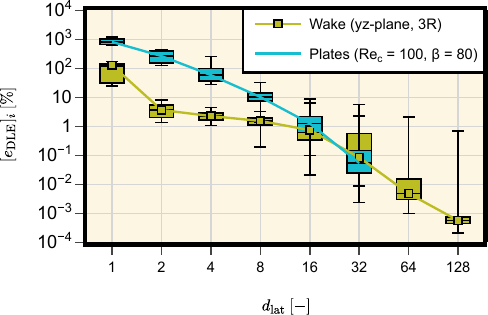}
    \caption{\textbf{DLE metric for different numbers of latent dimensions.} The boxes represent the lower and upper quartiles of the DLE values for a certain number of latent dimensions. The line plot shows the respective median. Each box and median is computed from combined DLE values for six decoders (from six-fold cross-validation). The autoencoders were trained on the wind turbine wake dataset (yz-plane subset, $\coordX=3R$) and the collinear plates case with $\mathrm{Re}_{\mathrm{c}} = 100$ and inflow angle $\beta = 80^{\circ}$.}
    \label{fig:NLM_Comparison}
\end{figure}

This argumentation is supported in Figure~\ref{fig:NLM_Comparison}, which shows the DLE for the wind turbine wake case in comparison to the second collinear plates case for different $d_{\mathrm{lat}}$ at the downstream position $\coordX=3R$ (yz-plane subset). As $d_{\mathrm{lat}}$ increases and thus compression decreases, the decoders exhibit more linear mapping, observable in both datasets. However, the DLE associated with the second reference dataset is much larger in the high compression range (i.e., a small $d_{\mathrm{lat}}$). This observation is consistent with the large differences that the LTR predicted between these two datasets (compare Figures~\ref{fig:LPOD_downstream}a and~\ref{fig:LPOD_reference_cases}b).

\section{Discussion}
\label{sec:discussion}

Both POD and the autoencoder extract coherent patterns from the data. For POD, these spatial patterns are contained in the POD modes, whilst the temporal evolution is encoded in the modal amplitudes. The autoencoder does not output the patterns in an explicit form; rather, the patterns are contained within the mapping functions of the encoder and decoder, $f_{\parameterSetEncoder}$ and $g_{\parameterSetDecoder}(\latentVariableVecTime{t})$, with the latent coordinates $\latentVariableVecTime{t}$ governing the specific realisation.

To assess whether the extracted spatial patterns generalise to be valid for the ground-truth turbulent flow (i.e., the turbulent flow observed over an ``infinite'' period of time), the dimension reduction methods were validated on unseen test datasets. If this was not the case, the model could learn a compression scheme that is specific to the considered turbulent flow segment and therefore not be capable of extracting generalisable patterns that also extend to unseen data. When used in the following, the term ``compression'' thus refers to the \emph{out-of-sample} compression on an unseen test dataset.

The effort to characterise the \emph{out-of-sample} compression potential of autoencoders addresses a broader methodological concern in the scientific machine learning literature. Recent evidence suggests that such machine learning studies are prone to reporting biases that arise because baseline methods are weak or no baseline method is considered at all, leading to a systematic overestimation of potential benefits~\cite{McGreivy2024}. For this reason, all autoencoder results are reported against POD as a baseline method using the exact same training-validation-test split with temporal gaps enforced between the subsets to minimise temporal leakage. Furthermore, the evaluation is restricted to the mean squared error as the loss function, so as not to create an artificial disadvantage for POD.

Under these conditions, POD also bounds the latent dimensions for which a non-linear advantage is possible. Because the snapshots are coherent fields rather than random vectors in $\mathbb{R}^{3\numberOfGridPoints}$, a reduced set of modes suffices to span them. POD therefore achieves an exact representation of the snapshots used to compute its modes ($1 - e_{\mathrm{NMSE}} = 100\,\%$) before $d_{\mathrm{lat}}$ reaches $3\numberOfGridPoints$; beyond this point no residual remains for a non-linear mapping to capture. At smaller latent dimensions, where compression necessarily discards information, POD is optimal among all linear decomposition methods\footnote{with respect to the NMSE}, but it is restricted to linear mappings. Since linear mappings are a special case of non-linear mappings, the POD curve can be interpreted as a lower bound for what general non-linear dimension reduction methods could achieve in this theoretical consideration\footnote{Most common CAE architectures, including the one proposed here, do not contain POD as a strict special case because of the convolutional kernel's local receptive field. Instead, CAEs require multiple successive convolutions to capture long-range spatial correlations.}.

The main challenge for non-linear dimension reduction methods lies in leveraging this theoretical potential. While POD exploits its linear reconstruction capabilities, the autoencoder performance in practical application depends on the neural network architecture that limits the representational power of the mapping functions, $f_{\parameterSetEncoder} \in \mathcal{F}$ and $g_{\parameterSetDecoder} \in \mathcal{G}$, and it depends on finding the optimal trainable parameters ($\parameterSetEncoder$ and $\parameterSetDecoder$). Due to randomised weight initialisation and stochastic optimisation algorithms (e.g., Adam, which is used here), even the best architectures identified through parameter studies are approximations of the theoretical optimum.

This practical limit for non-linear autoencoders is apparent in the first collinear plates case for strong compression ($d_{\mathrm{lat}} = 2$ and $d_{\mathrm{lat}} = 4$). Here, the CAE is not able to outperform the POD baseline. The most probable explanation for this case is a combination of insufficient representational power of the parametrised mapping functions and a non-optimal parameter optimisation that led to the CAE's compression limit. However, comparing the results for the other cases revealed another significant performance gap that is dependent on the dataset itself: while the autoencoder is able to outperform POD with a large margin in the reference cases, this is not possible in the wind turbine wake case.

These empirical findings are also supported in the literature. Autoencoders seem to outperform POD with a large margin for flows dominated by coherent structures with periodic dynamics, such as periodic vortex shedding forming a vortex street. Both \cite{Murata2019} and \cite{SoleraRico2024a} report substantial improvements over POD despite the use of additionally constrained variational autoencoders in the latter case. In contrast, the advantage of non-linear autoencoder-based decompositions appears considerably smaller for fully-turbulent flows, as reported, for example, in \cite{Fukami2021} for a turbulent channel flow.

The empirical observation of such an intrinsic compression limit for the non-linear autoencoder for different datasets motivates further discussion. In the following, first, the concept of an intrinsic compression limit for linear decomposition is therefore introduced. Then, a generalised intrinsic compression limit for non-linear dimension reduction methods is conceptualised to interpret the autoencoder's bounded reconstruction errors. Following that, the LTR is discussed as a practical method for a priori estimation of this intrinsic compression limit of a given dataset.

\paragraph{Linear intrinsic compression limit}

POD is able to represent the turbulent flow with a small number of modes whenever it consists of few TKE-dominant coherent spatial patterns whose temporal evolution can be superposed. This is often the case for flows with periodic vortex shedding (e.g., the \textsc{von Kármán} vortex street at low \textsc{Reynolds} numbers). In such cases, a sharp decline in the eigenvalue distribution can be observed, resulting in a low POD reconstruction error with few modes.

The concept of an intrinsic compression limit can thus be substantiated for the linear case:

\begin{equation}
    e_{\mathrm{NMSE}}(d_{\mathrm{lat}}) = 1 - \frac{\sum_{i=1}^{d_{\mathrm{lat}}} \lambda_{i}}{\sum_{i=1}^{3\numberOfGridPoints} \lambda_{i}} \, ,
\end{equation}

\noindent where the POD reconstruction error is formulated in terms of the eigenvalue distribution. Since POD is optimal in the mean-square sense, no other linear $d_{\mathrm{lat}}$-dimensional subspace can reduce this error further. The rate at which the eigenvalue distribution declines therefore governs the linear intrinsic compression limit. This linear intrinsic compression limit has been formalised, for example, in the \textsc{Kolmogorov} \emph{n-width}~\cite{Kolmogorov1936} as a theoretical lower bound on the worst-case error of a linear compressor for a given dataset. In practice, however, the eigenvalues are estimated from a finite snapshot ensemble, introducing an additional estimation error in the empirical covariance matrix.

\paragraph{Non-linear intrinsic compression limit}

For non-linear dimension reduction methods, defining an intrinsic compression limit is less straightforward than for linear methods, as non-linear couplings between spatial patterns are admissible. The best reconstruction error achievable using \emph{some} encoder--decoder pair can be reduced to arbitrarily small values~\cite{DeVore1989, Cohen2021}, but the resulting model could also become arbitrarily sensitive to small input perturbations. A physically meaningful, non-degenerate compression limit therefore additionally requires the encoder and decoder to be \textsc{Lipschitz} continuous, with a fixed bound on this sensitivity. This has been formalised in the \emph{approximation theory} literature as the \emph{non-linear $n$-width}, the non-linear analogue of the \textsc{Kolmogorov} \emph{n-width}~\cite{DeVore1989, Cohen2021}.

This constitutes a regularisation requirement that keeps the compression limit well defined but is also desirable for the autoencoder itself. A fixed bound on the amplification of small perturbations is a prerequisite for a meaningful model: without such a bound, two close data points could be mapped to distant points in latent space, or vice versa, so that interpolating or perturbing latent states would no longer produce physically consistent reconstructions. Meaningful physical interpolation is, however, an important requirement in the ROM context (e.g., parameter space interpolation in PS-ROM~\cite{Andersen2022}). Therefore, imposing the said regularisation requirement on the autoencoder is essential for the intended use of the latent representation.

Even an autoencoder that satisfies the regularisation requirement mentioned above can, at least in principle, achieve an arbitrarily small reconstruction error for a fixed latent dimension $d_{\mathrm{lat}}$, but only if the state of the turbulent flow were to evolve \emph{exactly} on a $d_{\mathrm{lat}}$-dimensional manifold that admits such a parametrisation. In practice, however, there is little reason to expect this idealisation to hold. First, there is no reason to assume that turbulent flow fields evolve on a manifold of this kind for an arbitrarily chosen $d_{\mathrm{lat}}$; the true \emph{intrinsic dimensionality} of the underlying dynamics is generally unknown. Second, even if such a manifold existed, nothing guarantees that a single and consistent global coordinate system can be found for it: a parametrisation that works well in a local neighbourhood does not need to extend across the whole manifold. For these reasons, a non-zero, relevant intrinsic compression limit must generally be expected for non-linear dimension reduction methods, just as it is for linear ones.

\paragraph{LTR metric}

The central problem in estimating the non-linear intrinsic compression limit therefore consists in determining whether the considered data points \emph{approximately} lie on a $d_{\mathrm{lat}}$-dimensional manifold that can be described with fewer coordinates, provided a suitable coordinate system is chosen. In this context, the POD modes are interpreted as the basis of a global orthonormal coordinate system. The problem of the intrinsic compression potential thus reduces to whether a more suitable local tangential coordinate system can be defined that describes the data with a lower error, or whether the orthonormal coordinate system is already optimal.

To address this problem, the LTR was introduced in the present work as an estimate: the nPOD basis defines the coordinate directions of a local tangential coordinate system, while the POD basis describes the directions of the global orthonormal coordinate system. Comparing the averaged reconstruction errors with respect to these two coordinate directions within the data points of a $k$-nearest neighbourhood defines the averaged local linearisation error. The minimum of this error under variation of the neighbourhood size is interpreted as an estimate of the non-linear intrinsic compression limit.

That this error exhibits a minimum at all follows from the way the data are split. As for the compression results, the LTR is evaluated \emph{out-of-sample}: the dataset is divided into training and validation subsets \emph{before} the neighbourhoods are defined. This ordering makes the overfitting point of the metric accessible, that is, the point at which a further reduction of the neighbourhood radius causes the local geometry of the training and validation points to diverge. Splitting the dataset \emph{afterwards}, within each individual neighbourhood, would remove this insight: the validation points would be restricted to lie close to the anchor points of the training subset, which constitutes a selection bias, and overlapping neighbourhoods would render the subsets non-distinct. The aim of modelling, however, is to identify generalisable structures within the dataset that occur in both distinct subsets.

Defined in this manner, the LTR can be used to predict the reconstruction performance of autoencoder-based ROMs a priori. Despite the inherent limitations of the metric, (1) the extrapolation from local observations to the reconstruction error of the autoencoder on the full dataset and (2) the reduced \emph{signal-to-noise ratio} associated with the local analysis, the LTR metric is able to reproduce the same trends as the observed CAE reconstruction errors. While the LTR captures the correct proportional trends, it does not reproduce the absolute values, which is attributable to the extrapolation from local to global reconstruction error.

As an a priori metric, the LTR is independent of the dimension reduction method and depends solely on the dataset: on the number of snapshots, the sampling rate and the spatial resolution. Its only parameter is the desired number of latent dimensions. The dataset specifications required for a given reconstruction performance can therefore be estimated before an autoencoder is developed. A higher sampling rate or a larger number of snapshots can resolve local structures that were previously obscured by undersampling, and the LTR responds to this change. The estimation is iterative: the LTR is evaluated on differently sampled datasets and the resulting trends are compared.

\paragraph{DLE metric}

The DLE confirms the LTR predictions: for the datasets under investigation, a large predicted non-linear potential leads to a pronounced non-linearity in the trained decoder. The DLE also permits these non-linearities to be interpreted, since a linear approximation represents non-linear dynamics based on dipoles. Because the DLE is evaluated separately for each latent dimension, it reveals which latent dimension encodes which flow feature, and whether that encoding exploits the non-linearity of the autoencoder.

The wind turbine wake illustrates this. In the near wake, the broad distribution of DLE values indicates that the non-linearity is carried by a small number of latent dimensions; this dominant non-linearity was already attributed to the motion of the hub vortex (see Section~\ref{sec:results:dle}). Further downstream, the DLE distribution narrows and its median decreases, indicating that no single feature dominates the encoding any longer. Accordingly, the performance of CAE and POD is comparable in this region of the flow.

That the non-linear dynamics are confined to a few latent dimensions in the near wake also points towards hybrid architectures that combine a CAE with POD (e.g., \cite{Somasekharan2025}): in a hybrid approach, the autoencoder would encode the dominant non-linear dynamics, while POD represents the remainder.

\paragraph{Computational costs}

\begin{figure}[tb]
    \centering
    \includegraphics{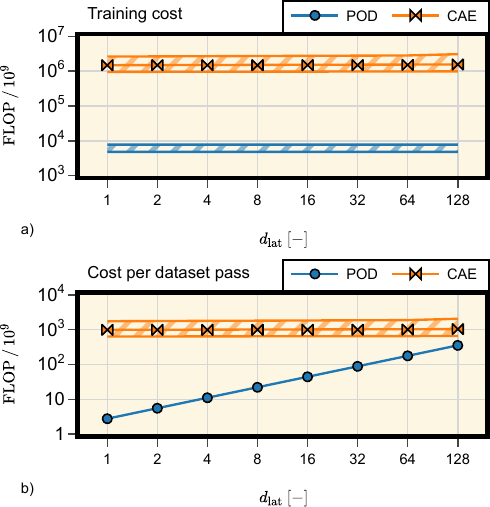}
    \caption{\textbf{Order-of-magnitude estimation of computational cost.} Costs are given in GFLOP for POD and CAE, considering $T = 100 \, 000$ snapshots and $n_{\mathrm{epochs}} = 500$. For the CAE, the lower and upper bounds correspond to the smallest and the largest network in the search space (see Table~\ref{tab:trainable_parameters}). For the POD computation in (a), the lower and upper bounds are derived in~\ref{app:computational_cost_estimation}. \textbf{(a)} CAE training or POD computation. \textbf{(b)} Dataset forward pass for both methods.}
    \label{fig:computational_cost}
\end{figure}

To conclude the discussion, the POD and CAE performance are further contextualised in terms of computational costs. Figure~\ref{fig:computational_cost} presents both (a) the costs for the training process or the POD computation and (b) the costs for a dataset \emph{forward pass} for each method. Here, the term dataset \emph{forward pass} refers to the encoding and subsequent decoding of the dataset. These costs are detailed in terms of \emph{floating-point operations} (FLOPs), where each multiplication and addition pair contributes two FLOPs. The cost estimation is performed using a dataset size $T = 100 \, 000$ and assuming $n_{\mathrm{epochs}} = 500$ training epochs. The cost estimation procedures for both methods are outlined in~\ref{app:computational_cost_estimation}.

The main trend observed in Figure~\ref{fig:computational_cost} is that POD requires several orders of magnitude fewer FLOPs in computational costs compared to the CAE training process in this estimation. However, the spatial resolution of the dataset ($48 \times 48$) is low, and the classical POD formulation is therefore cost-effective\footnote{Note that no temporal subsampling was applied in this work. Had $T$ been reduced below $3\numberOfGridPoints$, the method of snapshots would have become the cheaper formulation. This would have further lowered the POD cost reported here, provided the modes still converge on the subsampled dataset.}.

In both cases, the costs for training or POD computation remain almost constant for different numbers of latent dimensions. For the CAE, this is due to the static architecture of the convolutional layers, which do not change depending on the number of latent dimensions. Only the size of the considerably smaller dense subnetwork in the bottleneck changes depending on the number of latent dimensions. In the case of POD, the cost of accumulating the covariance matrix, which is independent of the number of latent dimensions, dominates to such an extent that any potential influence of partial eigensolvers for calculating the leading eigenvectors remains limited in this example.

The situation is somewhat more nuanced in the case of the dataset forward pass, as the cost of matrix multiplications for POD increases with the number of modes required. For the CAE, on the other hand, the cost remains largely constant again, as the architecture remains static except for the dense bottleneck subnetwork.

In the broader context, however, it must be noted that the largest cost associated with non-linear dimension reduction arises from the hyperparameter studies. These comprised 200 training runs (see~\ref{app:hyperparameter_optimisation}), in addition to preliminary studies on the selection of hyperparameters. The actual CAE cost incurred therefore exceeded the single-run values reported in Figure~\ref{fig:computational_cost}a and constituted the dominant computational overhead in this work.

As an order-of-magnitude estimate, the total computational cost for computing the LTR metric across all neighbourhoods is comparable to that of only a few full POD computations. This is because each nPOD is computed using the method of snapshots, which is the cheaper formulation in this case since a neighbourhood contains far fewer snapshots than $3\numberOfGridPoints$. Its cost then scales quadratically with the number of training snapshots contained in that neighbourhood: small neighbourhoods are therefore computationally cheap, and the overall cost is dominated by the comparatively few large neighbourhoods. Moreover, since large neighbourhoods necessarily overlap substantially, only a small number of them need to be evaluated to characterise the dataset. The exact computational cost naturally depends on the chosen size and number of neighbourhoods.

\section{Conclusion}
\label{sec:conclusion}

This work aimed to understand the potential benefit of autoencoders over POD for dimension reduction in turbulent flow regimes. Such a compression potential can be explained by the local geometry of the data points: POD achieves compact latent representations when TKE-dominant turbulent flow dynamics can be represented within few orthogonal modes. However, when a local tangential coordinate system describes the data points with lower reconstruction error, there is potential for autoencoders, or other non-linear dimension reduction methods, to outperform POD.

The LTR is introduced as an a priori estimator for this intrinsic compression potential. It predicts the dominant trends in the autoencoder performance gain over POD for all three investigated datasets. Besides the desired number of latent dimensions, the metric is free of parameters and requires no threshold to count the number of dominant directions. However, the predictive power of the LTR is limited by the elevated noise-to-signal ratio inherent to local as opposed to global analysis and by the need to extrapolate from the averaged local reconstruction error to the overall compression performance.

In addition, we present the DLE as a complementary a posteriori metric to quantify the non-linearity of the trained decoder based on a comparison with a linearised surrogate model. The DLE provides another perspective on the same observation: datasets for which the LTR indicates a high non-linear potential lead to trained decoders with pronounced non-linearity. Since the DLE requires a trained model, it verifies a posteriori what the LTR predicts beforehand.

At low computational cost, the LTR thus indicates whether the substantial effort of an autoencoder-based approach is justified for a given dataset or whether the well-established POD already provides a suitable latent representation, before any architecture search or hyperparameter tuning is undertaken. For practitioners that are interested in estimating the number of simulation timesteps or experimental measurements beforehand, the LTR can be applied in an iterative manner with a dataset that is enlarged in small steps. This holds because the LTR depends on the temporal resolution of the given dataset.

\section*{CRediT authorship contribution statement}
\noindent\textbf{Niklas Alexander Pissarski:} Conceptualization, Data curation, Formal analysis,
Investigation, Methodology, Software, Validation, Visualization, Writing -- original draft,
Writing -- review \& editing.

\noindent\textbf{Juan Pablo Murcia Leon:} Conceptualization, Supervision, Validation,
Writing -- review \& editing.

\noindent\textbf{S{\o}ren Juhl Andersen:} Conceptualization, Data curation, Funding acquisition,
Project administration, Resources, Supervision, Validation, Writing -- review \& editing.

\section*{Declaration of competing interest}

The authors declare that they have no known competing financial interests or personal relationships that could have appeared to influence the work reported in this paper.

\section*{Acknowledgements}
Computational resources have been provided by the DTU cluster Sophia~\cite{Sophia} and by the DTU Computing Center~\cite{Gbar}.

\section*{Funding}
This work was partially funded by the European Union through the MERIDIONAL project (\url{https://meridional.eu}; Horizon Europe, Grant Agreement No.~101084216), the DIGIWIND project (\url{https://digiwind.org}; Digital Europe Programme, Grant Agreement No.~101122836) and the INF4INiTY project (\url{https://inf4inity.com}; Horizon Europe, Grant Agreement No.~101136087).

\section*{Code availability}
The source code for computing the LTR and DLE and for training the autoencoder models is openly available under the MIT licence on GitLab at \url{https://gitlab.windenergy.dtu.dk/nalpi/how-nonlinear-is-my-flow}~\cite{PoF_PissarskiCode2026}. The version corresponding to this publication is tagged as \texttt{v1.0.0}.

\section*{Data availability}
The two reference datasets of the collinear plates cases are openly available via Zenodo~\cite{SoleraRico2024}. A representative subset of the wind turbine wake dataset, comprising the xy-plane, the xz-plane and the yz-plane at $\coordX=3R$, is openly available through DTU Data at \url{https://doi.org/10.11583/DTU.34044576}~\cite{PoF_PissarskiData2026}, together with the trained models. Additional data can be made available upon reasonable request to the corresponding author.

\section*{Declaration of generative AI and AI-assisted technologies in the manuscript preparation process}
During the preparation of this work, the first author used Claude (Anthropic) to refactor and document the source code and to improve the structure and wording of the manuscript. Suggestions from the tool were adopted only after careful review. The refactored code was verified to ensure its correctness. After using this tool, all three authors reviewed and edited the content as needed and take full responsibility for the content of the published article.

\bibliographystyle{elsarticle-num-names} 
\bibliography{main}

\begin{thebibliography}{60}
\expandafter\ifx\csname natexlab\endcsname\relax\def\natexlab#1{#1}\fi
\providecommand{\url}[1]{\texttt{#1}}
\providecommand{\href}[2]{#2}
\providecommand{\path}[1]{#1}
\providecommand{\DOIprefix}{doi:}
\providecommand{\ArXivprefix}{arXiv:}
\providecommand{\URLprefix}{URL: }
\providecommand{\Pubmedprefix}{pmid:}
\providecommand{\doi}[1]{\href{http://dx.doi.org/#1}{\path{#1}}}
\providecommand{\Pubmed}[1]{\href{pmid:#1}{\path{#1}}}
\providecommand{\bibinfo}[2]{#2}
\ifx\xfnm\relax \def\xfnm[#1]{\unskip,\space#1}\fi
\bibitem[{Yang et~al.(2024)Yang, Zhang, Abkar, and Anderson}]{Yang2024}
\bibinfo{author}{X.~Yang}, \bibinfo{author}{W.~Zhang}, \bibinfo{author}{M.~Abkar}, \bibinfo{author}{W.~Anderson},
\newblock \bibinfo{title}{Computational fluid dynamics: Its carbon footprint and role in carbon reduction},
\newblock \bibinfo{journal}{Journal of Renewable and Sustainable Energy} \bibinfo{volume}{16} (\bibinfo{year}{2024}). \DOIprefix\doi{10.1063/5.0217320}.
\bibitem[{Meneveau(2019)}]{Meneveau2019}
\bibinfo{author}{C.~Meneveau},
\newblock \bibinfo{title}{Big wind power: seven questions for turbulence research},
\newblock \bibinfo{journal}{Journal of Turbulence} \bibinfo{volume}{20} (\bibinfo{year}{2019}) \bibinfo{pages}{2--20}. \DOIprefix\doi{10.1080/14685248.2019.1584664}.
\bibitem[{Lumley(1967)}]{Lumley1967}
\bibinfo{author}{J.~L. Lumley},
\newblock \bibinfo{title}{{T}he {S}tructure of {I}nhomogeneous {T}urbulent {F}lows},
\newblock in: \bibinfo{editor}{A.~M. Yaglom}, \bibinfo{editor}{V.~I. Tatarskii} (Eds.), \bibinfo{booktitle}{Atmospheric Turbulence and Radio Wave Propagation}, \bibinfo{publisher}{Nauka Publishing House}, \bibinfo{address}{Moscow}, \bibinfo{year}{1967}, pp. \bibinfo{pages}{166--177}.
\bibitem[{Sirovich(1987)}]{Sirovich1987}
\bibinfo{author}{L.~Sirovich},
\newblock \bibinfo{title}{{T}urbulence and the {D}ynamics of {C}oherent {S}tructures, {P}art 1: {C}oherent {S}tructures},
\newblock \bibinfo{journal}{Quarterly of Applied Mathematics} \bibinfo{volume}{45} (\bibinfo{year}{1987}) \bibinfo{pages}{561--571}.
\bibitem[{Berkooz et~al.(1993)Berkooz, Holmes, and Lumley}]{Berkooz1993}
\bibinfo{author}{G.~Berkooz}, \bibinfo{author}{P.~Holmes}, \bibinfo{author}{J.~L. Lumley},
\newblock \bibinfo{title}{{T}he {P}roper {O}rthogonal {D}ecomposition in the {A}nalysis of {T}urbulent {F}lows},
\newblock \bibinfo{journal}{Annual Review of Fluid Mechanics} \bibinfo{volume}{25} (\bibinfo{year}{1993}) \bibinfo{pages}{539--575}. \DOIprefix\doi{10.1146/annurev.fl.25.010193.002543}.
\bibitem[{Milano and Koumoutsakos(2002)}]{Milano2002}
\bibinfo{author}{M.~Milano}, \bibinfo{author}{P.~Koumoutsakos},
\newblock \bibinfo{title}{Neural network modeling for near wall turbulent flow},
\newblock \bibinfo{journal}{Journal of Computational Physics} \bibinfo{volume}{182} (\bibinfo{year}{2002}) \bibinfo{pages}{1--26}. \DOIprefix\doi{10.1006/jcph.2002.7146}.
\bibitem[{Masci et~al.(2011)Masci, Meier, Cireşan, and Schmidhuber}]{Masci2011}
\bibinfo{author}{J.~Masci}, \bibinfo{author}{U.~Meier}, \bibinfo{author}{D.~Cireşan}, \bibinfo{author}{J.~Schmidhuber}, \bibinfo{title}{{S}tacked {C}onvolutional {A}uto-{E}ncoders for {H}ierarchical {F}eature {E}xtraction}, \bibinfo{publisher}{Springer Berlin Heidelberg}, \bibinfo{year}{2011}, pp. \bibinfo{pages}{52--59}. \DOIprefix\doi{10.1007/978-3-642-21735-7_7}.
\bibitem[{Morimoto et~al.(2021)Morimoto, Fukami, Zhang, Nair, and Fukagata}]{Morimoto2021}
\bibinfo{author}{M.~Morimoto}, \bibinfo{author}{K.~Fukami}, \bibinfo{author}{K.~Zhang}, \bibinfo{author}{A.~G. Nair}, \bibinfo{author}{K.~Fukagata},
\newblock \bibinfo{title}{Convolutional neural networks for fluid flow analysis: toward effective metamodeling and low dimensionalization},
\newblock \bibinfo{journal}{Theoretical and Computational Fluid Dynamics} \bibinfo{volume}{35} (\bibinfo{year}{2021}) \bibinfo{pages}{633--658}. \DOIprefix\doi{10.1007/s00162-021-00580-0}.
\bibitem[{Aubry et~al.(1988)Aubry, Holmes, Lumley, and Stone}]{Aubry1988}
\bibinfo{author}{N.~Aubry}, \bibinfo{author}{P.~Holmes}, \bibinfo{author}{J.~L. Lumley}, \bibinfo{author}{E.~Stone},
\newblock \bibinfo{title}{The dynamics of coherent structures in the wall region of a turbulent boundary layer},
\newblock \bibinfo{journal}{Journal of Fluid Mechanics} \bibinfo{volume}{192} (\bibinfo{year}{1988}) \bibinfo{pages}{115--173}. \DOIprefix\doi{10.1017/s0022112088001818}.
\bibitem[{Andersen and Murcia~Leon(2022)}]{Andersen2022}
\bibinfo{author}{S.~J. Andersen}, \bibinfo{author}{J.~P. Murcia~Leon},
\newblock \bibinfo{title}{Predictive and stochastic reduced-order modeling of wind turbine wake dynamics},
\newblock \bibinfo{journal}{Wind Energy Science} \bibinfo{volume}{7} (\bibinfo{year}{2022}) \bibinfo{pages}{2117--2133}. \DOIprefix\doi{10.5194/wes-7-2117-2022}.
\bibitem[{Murata et~al.(2019)Murata, Fukami, and Fukagata}]{Murata2019}
\bibinfo{author}{T.~Murata}, \bibinfo{author}{K.~Fukami}, \bibinfo{author}{K.~Fukagata},
\newblock \bibinfo{title}{Nonlinear mode decomposition with convolutional neural networks for fluid dynamics},
\newblock \bibinfo{journal}{Journal of Fluid Mechanics} \bibinfo{volume}{882} (\bibinfo{year}{2019}). \DOIprefix\doi{10.1017/jfm.2019.822}.
\bibitem[{Eivazi et~al.(2022)Eivazi, Le~Clainche, Hoyas, and Vinuesa}]{Eivazi2022}
\bibinfo{author}{H.~Eivazi}, \bibinfo{author}{S.~Le~Clainche}, \bibinfo{author}{S.~Hoyas}, \bibinfo{author}{R.~Vinuesa},
\newblock \bibinfo{title}{Towards extraction of orthogonal and parsimonious non-linear modes from turbulent flows},
\newblock \bibinfo{journal}{Expert Systems with Applications} \bibinfo{volume}{202} (\bibinfo{year}{2022}) \bibinfo{pages}{117038}. \DOIprefix\doi{10.1016/j.eswa.2022.117038}.
\bibitem[{Fukami et~al.(2020)Fukami, Nakamura, and Fukagata}]{Fukami2020}
\bibinfo{author}{K.~Fukami}, \bibinfo{author}{T.~Nakamura}, \bibinfo{author}{K.~Fukagata},
\newblock \bibinfo{title}{Convolutional neural network based hierarchical autoencoder for nonlinear mode decomposition of fluid field data},
\newblock \bibinfo{journal}{Physics of Fluids} \bibinfo{volume}{32} (\bibinfo{year}{2020}). \DOIprefix\doi{10.1063/5.0020721}.
\bibitem[{Doan et~al.(2023)Doan, Racca, and Magri}]{Doan2023}
\bibinfo{author}{N.~A.~K. Doan}, \bibinfo{author}{A.~Racca}, \bibinfo{author}{L.~Magri}, \bibinfo{title}{{C}onvolutional {A}utoencoder for the {S}patiotemporal {L}atent {R}epresentation of {T}urbulence}, \bibinfo{publisher}{Springer Nature Switzerland}, \bibinfo{year}{2023}, pp. \bibinfo{pages}{328--335}. \DOIprefix\doi{10.1007/978-3-031-36027-5\_24}.
\bibitem[{Solera-Rico et~al.(2024)Solera-Rico, Sanmiguel~Vila, Gómez-López, Wang, Almashjary, Dawson, and Vinuesa}]{SoleraRico2024a}
\bibinfo{author}{A.~Solera-Rico}, \bibinfo{author}{C.~Sanmiguel~Vila}, \bibinfo{author}{M.~Gómez-López}, \bibinfo{author}{Y.~Wang}, \bibinfo{author}{A.~Almashjary}, \bibinfo{author}{S.~T.~M. Dawson}, \bibinfo{author}{R.~Vinuesa},
\newblock \bibinfo{title}{$\beta$-{V}ariational autoencoders and transformers for reduced-order modelling of fluid flows},
\newblock \bibinfo{journal}{Nature Communications} \bibinfo{volume}{15} (\bibinfo{year}{2024}). \DOIprefix\doi{10.1038/s41467-024-45578-4}.
\bibitem[{Fukami et~al.(2021)Fukami, Hasegawa, Nakamura, Morimoto, and Fukagata}]{Fukami2021}
\bibinfo{author}{K.~Fukami}, \bibinfo{author}{K.~Hasegawa}, \bibinfo{author}{T.~Nakamura}, \bibinfo{author}{M.~Morimoto}, \bibinfo{author}{K.~Fukagata},
\newblock \bibinfo{title}{{M}odel {O}rder {R}eduction with {N}eural {N}etworks: {A}pplication to {L}aminar and {T}urbulent {F}lows},
\newblock \bibinfo{journal}{SN Computer Science} \bibinfo{volume}{2} (\bibinfo{year}{2021}). \DOIprefix\doi{10.1007/s42979-021-00867-3}.
\bibitem[{Binnie et~al.(2025)Binnie, Dłotko, Harvey, Malinowski, and Yim}]{Binnie2025}
\bibinfo{author}{J.~A.~D. Binnie}, \bibinfo{author}{P.~Dłotko}, \bibinfo{author}{J.~Harvey}, \bibinfo{author}{J.~Malinowski}, \bibinfo{author}{K.~M. Yim},
\newblock \bibinfo{title}{{A} {S}urvey of {D}imension {E}stimation {M}ethods},
\newblock \bibinfo{journal}{arXiv}  (\bibinfo{year}{2025}). \DOIprefix\doi{10.48550/ARXIV.2507.13887}. \href{http://arxiv.org/abs/2507.13887}{{\tt arXiv:2507.13887}}.
\bibitem[{Fan et~al.(2010)Fan, Gu, Qiao, and Zhang}]{Fan2010}
\bibinfo{author}{M.~Fan}, \bibinfo{author}{N.~Gu}, \bibinfo{author}{H.~Qiao}, \bibinfo{author}{B.~Zhang},
\newblock \bibinfo{title}{Intrinsic dimension estimation of data by principal component analysis},
\newblock \bibinfo{journal}{arXiv}  (\bibinfo{year}{2010}). \DOIprefix\doi{10.48550/ARXIV.1002.2050}. \href{http://arxiv.org/abs/1002.2050}{{\tt arXiv:1002.2050}}.
\bibitem[{Fukunaga and Olsen(1971)}]{Fukunaga1971}
\bibinfo{author}{K.~Fukunaga}, \bibinfo{author}{D.~R. Olsen},
\newblock \bibinfo{title}{{A}n {A}lgorithm for {F}inding {I}ntrinsic {D}imensionality of {D}ata},
\newblock \bibinfo{journal}{IEEE Transactions on Computers} \bibinfo{volume}{C-20} (\bibinfo{year}{1971}) \bibinfo{pages}{176--183}. \DOIprefix\doi{10.1109/t-c.1971.223208}.
\bibitem[{Lin and Zha(2008)}]{Lin2008}
\bibinfo{author}{T.~Lin}, \bibinfo{author}{H.~Zha},
\newblock \bibinfo{title}{{R}iemannian {M}anifold {L}earning},
\newblock \bibinfo{journal}{IEEE Transactions on Pattern Analysis and Machine Intelligence} \bibinfo{volume}{30} (\bibinfo{year}{2008}) \bibinfo{pages}{796--809}. \DOIprefix\doi{10.1109/tpami.2007.70735}.
\bibitem[{Kaiser(1960)}]{Kaiser1960}
\bibinfo{author}{H.~F. Kaiser},
\newblock \bibinfo{title}{{T}he {A}pplication of {E}lectronic {C}omputers to {F}actor {A}nalysis},
\newblock \bibinfo{journal}{Educational and Psychological Measurement} \bibinfo{volume}{20} (\bibinfo{year}{1960}) \bibinfo{pages}{141--151}. \DOIprefix\doi{10.1177/001316446002000116}.
\bibitem[{Jolliffe(2004)}]{Jolliffe2004}
\bibinfo{author}{I.~T. Jolliffe}, \bibinfo{title}{Principal component analysis}, Springer series in statistics, \bibinfo{publisher}{Springer}, \bibinfo{address}{New York}, \bibinfo{year}{2004}.
\bibitem[{Le~Cun(1989)}]{LeCun1989}
\bibinfo{author}{Y.~Le~Cun}, \bibinfo{title}{{G}eneralization and {N}etwork {D}esign {S}trategies}, \bibinfo{type}{techreport} \bibinfo{number}{CRG-TR-89-4}, Department of Computer Science, University of Toronto, \bibinfo{year}{1989}.
\bibitem[{Baldi and Hornik(1989)}]{Baldi1989}
\bibinfo{author}{P.~Baldi}, \bibinfo{author}{K.~Hornik},
\newblock \bibinfo{title}{Neural networks and principal component analysis: Learning from examples without local minima},
\newblock \bibinfo{journal}{Neural Networks} \bibinfo{volume}{2} (\bibinfo{year}{1989}) \bibinfo{pages}{53--58}. \DOIprefix\doi{10.1016/0893-6080(89)90014-2}.
\bibitem[{Ioffe and Szegedy(2015)}]{Ioffe2015}
\bibinfo{author}{S.~Ioffe}, \bibinfo{author}{C.~Szegedy},
\newblock \bibinfo{title}{{B}atch {N}ormalization: {A}ccelerating {D}eep {N}etwork {T}raining by {R}educing {I}nternal {C}ovariate {S}hift},
\newblock in: \bibinfo{editor}{F.~Bach}, \bibinfo{editor}{D.~Blei} (Eds.), \bibinfo{booktitle}{Proceedings of the 32nd International Conference on Machine Learning}, volume~\bibinfo{volume}{37} of \textit{\bibinfo{series}{Proceedings of Machine Learning Research}}, \bibinfo{publisher}{PMLR}, \bibinfo{address}{Lille, France}, \bibinfo{year}{2015}, pp. \bibinfo{pages}{448--456}. \URLprefix \url{https://proceedings.mlr.press/v37/ioffe15.html}.
\bibitem[{He et~al.(2015)He, Zhang, Ren, and Sun}]{He2015}
\bibinfo{author}{K.~He}, \bibinfo{author}{X.~Zhang}, \bibinfo{author}{S.~Ren}, \bibinfo{author}{J.~Sun},
\newblock \bibinfo{title}{{D}elving {D}eep into {R}ectifiers: {S}urpassing {H}uman-{L}evel {P}erformance on {I}mage{N}et {C}lassification},
\newblock in: \bibinfo{booktitle}{2015 IEEE International Conference on Computer Vision (ICCV)}, \bibinfo{publisher}{IEEE}, \bibinfo{year}{2015}, pp. \bibinfo{pages}{1026--1034}. \DOIprefix\doi{10.1109/iccv.2015.123}.
\bibitem[{Saxe et~al.(2013)Saxe, McClelland, and Ganguli}]{Saxe2013}
\bibinfo{author}{A.~M. Saxe}, \bibinfo{author}{J.~L. McClelland}, \bibinfo{author}{S.~Ganguli},
\newblock \bibinfo{title}{Exact solutions to the nonlinear dynamics of learning in deep linear neural networks},
\newblock \bibinfo{journal}{arXiv}  (\bibinfo{year}{2013}). \DOIprefix\doi{10.48550/ARXIV.1312.6120}. \href{http://arxiv.org/abs/1312.6120}{{\tt arXiv:1312.6120}}.
\bibitem[{Barron(2017)}]{Barron2017}
\bibinfo{author}{J.~T. Barron},
\newblock \bibinfo{title}{{C}ontinuously {D}ifferentiable {E}xponential {L}inear {U}nits},
\newblock \bibinfo{journal}{arXiv}  (\bibinfo{year}{2017}). \DOIprefix\doi{10.48550/ARXIV.1704.07483}. \href{http://arxiv.org/abs/1704.07483}{{\tt arXiv:1704.07483}}.
\bibitem[{Akiba et~al.(2019)Akiba, Sano, Yanase, Ohta, and Koyama}]{Akiba2019}
\bibinfo{author}{T.~Akiba}, \bibinfo{author}{S.~Sano}, \bibinfo{author}{T.~Yanase}, \bibinfo{author}{T.~Ohta}, \bibinfo{author}{M.~Koyama},
\newblock \bibinfo{title}{{O}ptuna: {A} {N}ext-generation {H}yperparameter {O}ptimization {F}ramework},
\newblock \bibinfo{journal}{arXiv}  (\bibinfo{year}{2019}). \DOIprefix\doi{10.48550/ARXIV.1907.10902}. \href{http://arxiv.org/abs/1907.10902}{{\tt arXiv:1907.10902}}.
\bibitem[{Paszke et~al.(2019)Paszke, Gross, Massa, Lerer, Bradbury, Chanan, Killeen, Lin, Gimelshein, Antiga, Desmaison, K\"{o}pf, Yang, DeVito, Raison, Tejani, Chilamkurthy, Steiner, Fang, Bai, and Chintala}]{Paszke2019}
\bibinfo{author}{A.~Paszke}, \bibinfo{author}{S.~Gross}, \bibinfo{author}{F.~Massa}, \bibinfo{author}{A.~Lerer}, \bibinfo{author}{J.~Bradbury}, \bibinfo{author}{G.~Chanan}, \bibinfo{author}{T.~Killeen}, \bibinfo{author}{Z.~Lin}, \bibinfo{author}{N.~Gimelshein}, \bibinfo{author}{L.~Antiga}, \bibinfo{author}{A.~Desmaison}, \bibinfo{author}{A.~K\"{o}pf}, \bibinfo{author}{E.~Yang}, \bibinfo{author}{Z.~DeVito}, \bibinfo{author}{M.~Raison}, \bibinfo{author}{A.~Tejani}, \bibinfo{author}{S.~Chilamkurthy}, \bibinfo{author}{B.~Steiner}, \bibinfo{author}{L.~Fang}, \bibinfo{author}{J.~Bai}, \bibinfo{author}{S.~Chintala},
\newblock \bibinfo{title}{Pytorch: an imperative style, high-performance deep learning library},
\newblock in: \bibinfo{booktitle}{Proceedings of the 33rd International Conference on Neural Information Processing Systems}, \bibinfo{publisher}{Curran Associates Inc.}, \bibinfo{address}{Red Hook, NY, USA}, \bibinfo{year}{2019}, pp. \bibinfo{pages}{8026--8037}.
\bibitem[{Kingma and Ba(2014)}]{Kingma2014}
\bibinfo{author}{D.~P. Kingma}, \bibinfo{author}{J.~Ba},
\newblock \bibinfo{title}{{A}dam: {A} {M}ethod for {S}tochastic {O}ptimization},
\newblock \bibinfo{journal}{arXiv}  (\bibinfo{year}{2014}). \DOIprefix\doi{10.48550/ARXIV.1412.6980}. \href{http://arxiv.org/abs/1412.6980}{{\tt arXiv:1412.6980}}.
\bibitem[{Noack et~al.(2003)Noack, Afanasiev, Morzyński, Tadmor, and Thiele}]{Noack2003}
\bibinfo{author}{B.~R. Noack}, \bibinfo{author}{K.~Afanasiev}, \bibinfo{author}{M.~Morzyński}, \bibinfo{author}{G.~Tadmor}, \bibinfo{author}{F.~Thiele},
\newblock \bibinfo{title}{A hierarchy of low-dimensional models for the transient and post-transient cylinder wake},
\newblock \bibinfo{journal}{Journal of Fluid Mechanics} \bibinfo{volume}{497} (\bibinfo{year}{2003}) \bibinfo{pages}{335--363}. \DOIprefix\doi{10.1017/s0022112003006694}.
\bibitem[{Asztalos et~al.(2023)Asztalos, Almashjary, and Dawson}]{Asztalos2023}
\bibinfo{author}{K.~J. Asztalos}, \bibinfo{author}{A.~Almashjary}, \bibinfo{author}{S.~T.~M. Dawson},
\newblock \bibinfo{title}{Galerkin spectral estimation of vortex-dominated wake flows},
\newblock \bibinfo{journal}{Theoretical and Computational Fluid Dynamics} \bibinfo{volume}{38} (\bibinfo{year}{2023}) \bibinfo{pages}{801--823}. \DOIprefix\doi{10.1007/s00162-023-00670-1}.
\bibitem[{Solera-Rico et~al.(2024)Solera-Rico, Sanmiguel~Vila, Gómez, Wang, Almashjary, Dawson, and Vinuesa}]{SoleraRico2024}
\bibinfo{author}{A.~Solera-Rico}, \bibinfo{author}{C.~Sanmiguel~Vila}, \bibinfo{author}{M.~A. Gómez}, \bibinfo{author}{Y.~Wang}, \bibinfo{author}{A.~Almashjary}, \bibinfo{author}{S.~Dawson}, \bibinfo{author}{R.~Vinuesa}, \bibinfo{title}{{D}ataset of "$\beta$-{V}ariational autoencoders and transformers for reduced-order modelling of fluid flows"}, \bibinfo{year}{2024}. \DOIprefix\doi{10.5281/ZENODO.10501216}.
\bibitem[{Michelsen(1992)}]{Michelsen1992}
\bibinfo{author}{J.~A. Michelsen}, \bibinfo{title}{{B}asis{3D} - a {P}latform for {D}evelopment of {M}ultiblock {PDE} {S}olvers}, \bibinfo{type}{resreport} \bibinfo{number}{AFM 92-05}, Technical University of Denmark, \bibinfo{year}{1992}.
\bibitem[{Michelsen(1994)}]{Michelsen1994}
\bibinfo{author}{J.~A. Michelsen}, \bibinfo{title}{{B}lock structured {M}ultigrid solution of {2D} and {3D} elliptic {PDE}'s}, \bibinfo{type}{resreport} \bibinfo{number}{AFM 94-06}, Technical University of Denmark, \bibinfo{year}{1994}.
\bibitem[{Sørensen(1995)}]{Soerensen1995}
\bibinfo{author}{N.~N. Sørensen}, \bibinfo{title}{{G}eneral {P}urpose {F}low {S}olver {A}pplied to {F}low over {H}ills}, \bibinfo{type}{techreport} \bibinfo{number}{Risø-R-827-(EN)}, Risø National Laboratory, Roskilde, \bibinfo{year}{1995}.
\bibitem[{Deardorff(1980)}]{Deardorff1980}
\bibinfo{author}{J.~W. Deardorff},
\newblock \bibinfo{title}{Stratocumulus-capped mixed layers derived from a three-dimensional model},
\newblock \bibinfo{journal}{Boundary-Layer Meteorology} \bibinfo{volume}{18} (\bibinfo{year}{1980}) \bibinfo{pages}{495--527}. \DOIprefix\doi{10.1007/bf00119502}.
\bibitem[{Shen et~al.(2003)Shen, Michelsen, Sørensen, and Nørkær~Sørensen}]{Shen2003}
\bibinfo{author}{W.~Z. Shen}, \bibinfo{author}{J.~A. Michelsen}, \bibinfo{author}{N.~N. Sørensen}, \bibinfo{author}{J.~Nørkær~Sørensen},
\newblock \bibinfo{title}{{A}n {I}mproved {SIMPLEC} {M}ethod for {S}teady and {U}nsteady {F}low {C}omputations},
\newblock \bibinfo{journal}{Numerical Heat Transfer, Part B: Fundamentals} \bibinfo{volume}{43} (\bibinfo{year}{2003}) \bibinfo{pages}{221--239}. \DOIprefix\doi{10.1080/713836202}.
\bibitem[{Rhie and Chow(1983)}]{Rhie1983}
\bibinfo{author}{C.~M. Rhie}, \bibinfo{author}{W.~L. Chow},
\newblock \bibinfo{title}{Numerical study of the turbulent flow past an airfoil with trailing edge separation},
\newblock \bibinfo{journal}{AIAA Journal} \bibinfo{volume}{21} (\bibinfo{year}{1983}) \bibinfo{pages}{1525--1532}. \DOIprefix\doi{10.2514/3.8284}.
\bibitem[{Gaertner et~al.(2020)Gaertner, Rinker, Sethuraman, Zahle, Anderson, Barter, Abbas, Meng, Bortolotti, Skrzypinski, Scott, Feil, Bredmose, Dykes, Shields, Allen, and Viselli}]{Gaertner2020}
\bibinfo{author}{E.~Gaertner}, \bibinfo{author}{J.~Rinker}, \bibinfo{author}{L.~Sethuraman}, \bibinfo{author}{F.~Zahle}, \bibinfo{author}{B.~Anderson}, \bibinfo{author}{G.~Barter}, \bibinfo{author}{N.~Abbas}, \bibinfo{author}{F.~Meng}, \bibinfo{author}{P.~Bortolotti}, \bibinfo{author}{W.~Skrzypinski}, \bibinfo{author}{G.~Scott}, \bibinfo{author}{R.~Feil}, \bibinfo{author}{H.~Bredmose}, \bibinfo{author}{K.~Dykes}, \bibinfo{author}{M.~Shields}, \bibinfo{author}{C.~Allen}, \bibinfo{author}{A.~Viselli}, \bibinfo{title}{{D}efinition of the {IEA} 15-{M}egawatt {O}ffshore {R}eference {W}ind}, \bibinfo{type}{techreport} \bibinfo{number}{NREL/TP-5000-75698}, National Renewable Energy Laboratory, \bibinfo{year}{2020}.
\bibitem[{Larsen and Hansen(2025)}]{Larsen2025}
\bibinfo{author}{T.~J. Larsen}, \bibinfo{author}{A.~M. Hansen}, \bibinfo{title}{{H}ow 2 {HAWC2}, the user's manual}, \bibinfo{type}{techreport} \bibinfo{number}{Risø-R-1597}, Risø National Laboratory, Roskilde, \bibinfo{year}{2025}.
\bibitem[{Mikkelsen(2004)}]{Mikkelsen2004}
\bibinfo{author}{R.~F. Mikkelsen}, \bibinfo{title}{{A}ctuator {D}isc {M}ethods {A}pplied to {W}ind {T}urbines}, \bibinfo{type}{phdthesis} \bibinfo{number}{MEK-FM-PHD 2003-02}, Technical University of Denmark, \bibinfo{year}{2004}.
\bibitem[{Hodgson et~al.(2021)Hodgson, Andersen, Troldborg, Forsting, Mikkelsen, and Sørensen}]{Hodgson2021}
\bibinfo{author}{E.~L. Hodgson}, \bibinfo{author}{S.~J. Andersen}, \bibinfo{author}{N.~Troldborg}, \bibinfo{author}{A.~M. Forsting}, \bibinfo{author}{R.~F. Mikkelsen}, \bibinfo{author}{J.~N. Sørensen},
\newblock \bibinfo{title}{{A} {Q}uantitative {C}omparison of {A}eroelastic {C}omputations using {Flex5} and {A}ctuator {M}ethods in {LES}},
\newblock \bibinfo{journal}{Journal of Physics: Conference Series} \bibinfo{volume}{1934} (\bibinfo{year}{2021}) \bibinfo{pages}{012014}. \DOIprefix\doi{10.1088/1742-6596/1934/1/012014}.
\bibitem[{Pedersen et~al.(2026)Pedersen, Steiner, Nilsen, Lohmann, Hodgson, Riva, Troldborg, Andersen, Larsen, Verelst et~al.}]{Pedersen2026}
\bibinfo{author}{M.~M. Pedersen}, \bibinfo{author}{J.~Steiner}, \bibinfo{author}{M.~B. Nilsen}, \bibinfo{author}{J.~Lohmann}, \bibinfo{author}{E.~L. Hodgson}, \bibinfo{author}{R.~Riva}, \bibinfo{author}{N.~Troldborg}, \bibinfo{author}{S.~J. Andersen}, \bibinfo{author}{G.~Larsen}, \bibinfo{author}{D.~R. Verelst}, et~al., \bibinfo{title}{{D}ynamiks 0.0.4: {A}n open-source {D}ynamic {W}ind {S}ystem {S}imulator}, \bibinfo{howpublished}{GitLab}, \bibinfo{year}{2026}.
\bibitem[{Yang and Sotiropoulos(2019)}]{Yang2019}
\bibinfo{author}{X.~Yang}, \bibinfo{author}{F.~Sotiropoulos},
\newblock \bibinfo{title}{{A} {R}eview on the {M}eandering of {W}ind {T}urbine {W}akes},
\newblock \bibinfo{journal}{Energies} \bibinfo{volume}{12} (\bibinfo{year}{2019}) \bibinfo{pages}{4725}. \DOIprefix\doi{10.3390/en12244725}.
\bibitem[{Sørensen et~al.(2015)Sørensen, Mikkelsen, Henningson, Ivanell, Sarmast, and Andersen}]{Soerensen2015}
\bibinfo{author}{J.~N. Sørensen}, \bibinfo{author}{R.~F. Mikkelsen}, \bibinfo{author}{D.~S. Henningson}, \bibinfo{author}{S.~Ivanell}, \bibinfo{author}{S.~Sarmast}, \bibinfo{author}{S.~J. Andersen},
\newblock \bibinfo{title}{Simulation of wind turbine wakes using the actuator line technique},
\newblock \bibinfo{journal}{Philosophical Transactions of the Royal Society A: Mathematical, Physical and Engineering Sciences} \bibinfo{volume}{373} (\bibinfo{year}{2015}) \bibinfo{pages}{20140071}. \DOIprefix\doi{10.1098/rsta.2014.0071}.
\bibitem[{McGreivy and Hakim(2024)}]{McGreivy2024}
\bibinfo{author}{N.~McGreivy}, \bibinfo{author}{A.~Hakim},
\newblock \bibinfo{title}{Weak baselines and reporting biases lead to overoptimism in machine learning for fluid-related partial differential equations},
\newblock \bibinfo{journal}{Nature Machine Intelligence} \bibinfo{volume}{6} (\bibinfo{year}{2024}) \bibinfo{pages}{1256--1269}. \DOIprefix\doi{10.1038/s42256-024-00897-5}.
\bibitem[{Kolmogorov(1936)}]{Kolmogorov1936}
\bibinfo{author}{A.~Kolmogorov},
\newblock \bibinfo{title}{{Ü}ber die beste {A}nnäherung von {F}unktionen einer gegebenen {F}unktionenklasse},
\newblock \bibinfo{journal}{The Annals of Mathematics} \bibinfo{volume}{37} (\bibinfo{year}{1936}) \bibinfo{pages}{107}. \DOIprefix\doi{10.2307/1968691}.
\bibitem[{DeVore et~al.(1989)DeVore, Howard, and Micchelli}]{DeVore1989}
\bibinfo{author}{R.~A. DeVore}, \bibinfo{author}{R.~Howard}, \bibinfo{author}{C.~Micchelli},
\newblock \bibinfo{title}{Optimal nonlinear approximation},
\newblock \bibinfo{journal}{Manuscripta Mathematica} \bibinfo{volume}{63} (\bibinfo{year}{1989}) \bibinfo{pages}{469--478}. \DOIprefix\doi{10.1007/bf01171759}.
\bibitem[{Cohen et~al.(2021)Cohen, DeVore, Petrova, and Wojtaszczyk}]{Cohen2021}
\bibinfo{author}{A.~Cohen}, \bibinfo{author}{R.~DeVore}, \bibinfo{author}{G.~Petrova}, \bibinfo{author}{P.~Wojtaszczyk},
\newblock \bibinfo{title}{{O}ptimal {S}table {N}onlinear {A}pproximation},
\newblock \bibinfo{journal}{Foundations of Computational Mathematics} \bibinfo{volume}{22} (\bibinfo{year}{2021}) \bibinfo{pages}{607--648}. \DOIprefix\doi{10.1007/s10208-021-09494-z}.
\bibitem[{Somasekharan and Pan(2025)}]{Somasekharan2025}
\bibinfo{author}{N.~Somasekharan}, \bibinfo{author}{S.~Pan},
\newblock \bibinfo{title}{{B}eyond the {K}olmogorov barrier: a learnable weighted hybrid autoencoder for model order reduction},
\newblock \bibinfo{journal}{Proceedings of the Royal Society A: Mathematical, Physical and Engineering Sciences} \bibinfo{volume}{481} (\bibinfo{year}{2025}). \DOIprefix\doi{10.1098/rspa.2025.0159}.
\bibitem[{{Technical University of Denmark}(2026)}]{Sophia}
\bibinfo{author}{{Technical University of Denmark}}, \bibinfo{title}{{S}ophia {HPC} {C}luster}, \bibinfo{year}{2026}. \URLprefix \url{https://dtu-sophia.github.io/docs/}. \DOIprefix\doi{10.57940/FAFC-6M81}.
\bibitem[{{DTU Computing Center}(2026)}]{Gbar}
\bibinfo{author}{{DTU Computing Center}}, \bibinfo{title}{{DTU Computing Center resources}}, \bibinfo{year}{2026}. \URLprefix \url{https://doi.org/10.48714/DTU.HPC.0001}. \DOIprefix\doi{10.48714/DTU.HPC.0001}.
\bibitem[{Pissarski et~al.(2026{\natexlab{a}})Pissarski, Murcia~Leon, and Andersen}]{PoF_PissarskiCode2026}
\bibinfo{author}{N.~A. Pissarski}, \bibinfo{author}{J.~P. Murcia~Leon}, \bibinfo{author}{S.~J. Andersen}, \bibinfo{title}{{C}ode for ''{A} {P}riori {E}stimation of the {N}on-{L}inear {C}ompression {P}otential and {Q}uantification of {A}utoencoder {N}on-{L}inearity in {T}urbulent {F}lows"}, \bibinfo{howpublished}{GitLab}, \bibinfo{year}{2026}{\natexlab{a}}. \URLprefix \url{https://gitlab.windenergy.dtu.dk/nalpi/how-nonlinear-is-my-flow}, \bibinfo{note}{software}.
\bibitem[{Pissarski et~al.(2026{\natexlab{b}})Pissarski, Murcia~Leon, and Andersen}]{PoF_PissarskiData2026}
\bibinfo{author}{N.~A. Pissarski}, \bibinfo{author}{J.~P. Murcia~Leon}, \bibinfo{author}{S.~J. Andersen}, \bibinfo{title}{{D}ata for ''{A} {P}riori {E}stimation of the {N}on-{L}inear {C}ompression {P}otential and {Q}uantification of {A}utoencoder {N}on-{L}inearity in {T}urbulent {F}lows"}, \bibinfo{howpublished}{DTU Data}, \bibinfo{year}{2026}{\natexlab{b}}. \URLprefix \url{https://doi.org/10.11583/DTU.34044576}. \DOIprefix\doi{10.11583/DTU.34044576}, \bibinfo{note}{dataset}.
\bibitem[{Lehoucq et~al.(1998)Lehoucq, Sorensen, and Yang}]{Lehoucq1998}
\bibinfo{author}{R.~B. Lehoucq}, \bibinfo{author}{D.~C. Sorensen}, \bibinfo{author}{C.~Yang}, \bibinfo{title}{{ARPACK} {U}sers’ {G}uide: {S}olution of {L}arge-{S}cale {E}igenvalue {P}roblems with {I}mplicitly {R}estarted {A}rnoldi {M}ethods}, \bibinfo{publisher}{Society for Industrial and Applied Mathematics}, \bibinfo{year}{1998}. \DOIprefix\doi{10.1137/1.9780898719628}.
\bibitem[{Golub and Loan(2013)}]{Golub2013}
\bibinfo{author}{G.~H. Golub}, \bibinfo{author}{C.~F.~V. Loan}, \bibinfo{title}{Matrix computations}, Johns Hopkins studies in the mathematical sciences, \bibinfo{edition}{4th} ed., \bibinfo{publisher}{Johns Hopkins Univ. Press}, \bibinfo{address}{Baltimore, MD}, \bibinfo{year}{2013}. \bibinfo{note}{In association with the Department of Mathematical Sciences, The Johns Hopkins University.}
\bibitem[{{Meta AI}(2026)}]{Fvcore2026}
\bibinfo{author}{{Meta AI}}, \bibinfo{title}{fvcore}, \bibinfo{howpublished}{GitHub}, \bibinfo{year}{2026}. \URLprefix \url{https://github.com/facebookresearch/fvcore}.
\bibitem[{Kaplan et~al.(2020)Kaplan, McCandlish, Henighan, Brown, Chess, Child, Gray, Radford, Wu, and Amodei}]{Kaplan2020}
\bibinfo{author}{J.~Kaplan}, \bibinfo{author}{S.~McCandlish}, \bibinfo{author}{T.~Henighan}, \bibinfo{author}{T.~B. Brown}, \bibinfo{author}{B.~Chess}, \bibinfo{author}{R.~Child}, \bibinfo{author}{S.~Gray}, \bibinfo{author}{A.~Radford}, \bibinfo{author}{J.~Wu}, \bibinfo{author}{D.~Amodei},
\newblock \bibinfo{title}{{S}caling {L}aws for {N}eural {L}anguage {M}odels},
\newblock \bibinfo{journal}{arXiv}  (\bibinfo{year}{2020}). \href{http://arxiv.org/abs/2001.08361}{{\tt arXiv:2001.08361}}.

\end{thebibliography}

\clearpage
\appendix
\section{Mathematical formulation of methods}
\label{app:mathematical_formulation}

\subsection{Convolutional layer}
\label{app:mathematical_formulation:convolutional_layer}

To form the CAE input, the snapshot $\stateMatrixTime{t} \in \mathbb{R}^{3\numberOfGridPoints}$ is first restructured into a feature tensor $\boldsymbol{F}^{[0]} \in \mathbb{R}^{3 \times \numberOfGridPointsIdx{1} \times \numberOfGridPointsIdx{2}}$ with three separate input feature maps for each fluctuating velocity component. $\numberOfGridPointsIdx{1}$ and $\numberOfGridPointsIdx{2}$ are the number of grid points in the snapshot plane. A convolutional layer then applies a set of trainable filters to the input feature map to generate the output feature maps $\boldsymbol{F}^{[l]} \in \mathbb{R}^{C_{l} \times \numberOfGridPointsIdx{1} \times \numberOfGridPointsIdx{2}}$. A single output feature map $c \in \{1, \, \ldots, \, C_{l}\}$ is computed through summing the 2D convolutions over all input feature maps $c^{\prime}$:

\begin{equation}
    \boldsymbol{F}_{c}^{[l]} = \varphi\Bigl( \, \sum_{c' = 1}^{C_{l-1}} \boldsymbol{W}^{[l]}_{c, \, c^{\prime}} * \boldsymbol{F}_{c^{\prime}}^{[l-1]} + \boldsymbol{b}_{c}^{[l]}\,\Bigr) \, ,
\end{equation}

\noindent where $\varphi(\cdot)$ is the activation function, $\boldsymbol{W}^{[l]}_{c, \, c^{\prime}} \in \mathbb{R}^{K_{1} \times K_{2}}$ is the 2D filter kernel that is applied to feature map $c^{\prime}$ and $\boldsymbol{b}_{c}^{[l]} \in \mathbb{R}$ is the bias. The operator $(*)$ denotes the 2D discrete convolution that is defined between a generic 2D filter kernel $\boldsymbol{W} \in \mathbb{R}^{K_{1} \times K_{2}}$ and a generic input $\boldsymbol{F} \in \mathbb{R}^{\numberOfGridPointsIdx{1} \times \numberOfGridPointsIdx{2}}$ for a single grid point $(p_{1}, \, p_{2}) \in \{1, \, \ldots, \, \numberOfGridPointsIdx{1}\} \times \{1, \, \ldots, \, \numberOfGridPointsIdx{2}\}$ as:

\begin{equation}
    \bigl( \boldsymbol{W} * \boldsymbol{F}\bigr)_{p_{1}, \, p_{2}} = \sum_{s_1 \in \Delta_{1}}\sum_{s_2 \in \Delta_{2}} \boldsymbol{W}_{s_{1}, \, s_{2}} \cdot \boldsymbol{F}_{p_{1} + s_{1}, \, p_{2} + s_{2}}
\end{equation}

\noindent with indices $\Delta_i = \bigl\{ -\left\lfloor K_i/2 \right\rfloor, \, \ldots, \, 0, \, \ldots, \, \left\lfloor K_i/2 \right\rfloor \bigr\}$ for $i \in \{1, \, 2\}$ and assuming an odd number of filter dimensions $K_{1}$ and $K_{2}$ (i.e., a centred kernel).

The kernel matrices $\boldsymbol{W}^{[l]}$ and the bias vectors $\boldsymbol{b}^{[l]}$ are part of the trainable parameter sets $\parameterSetEncoder$ and $\parameterSetDecoder$ of the neural network.

\subsection{Definition of the LTR}
\label{app:mathematical_formulation:ltr}

The procedure of computing the LTR can be formalised as follows. The local neighbourhood set $\mathcal{N} = \mathcal{N}_{\mathrm{train}} \cup \mathcal{N}_{\mathrm{valid}}$ contains the indices of the $k = k_{\mathrm{train}} + k_{\mathrm{valid}}$ snapshots that form the local neighbourhood of snapshots $\stateMatrix_{\mathcal{N}}$:

\begin{equation}
    \stateMatrix_{\mathcal{N}} = \bigl[\stateMatrixTime{t}\bigr]_{t \in \mathcal{N}} \in \mathbb{R}^{3\numberOfGridPoints \times k} \, .
\end{equation}

Instead of examining the high-dimensional snapshots $\stateMatrix_{\mathcal{N}}$, the LTR method can also be applied to a lower-dimensional representation $\latentVariableVecIdx{\mathcal{N}}$, provided that the dimensional reduction is linear.

In the proposed procedure, the neighbourhood points from the training subset are identified via the $k$-nearest neighbour method ($k$-NN), selecting those $k_{\mathrm{train}}$ snapshots from the training subset with the smallest \textsc{Euclidean} distances to an anchor point $\stateMatrixTime{a}$ with $a \in \mathcal{A}$:

\begin{equation}
    \mathcal{N}_{\mathrm{train}} = \Bigl\{t \in \mathcal{T}_{\mathrm{train}}: \bigl\|\stateMatrixTime{a} - \stateMatrixTime{t}\bigr\|_{2} \leq r\bigl(a, \, k_{\mathrm{train}}\bigr)\Bigr\} \, ,
\end{equation}

\noindent where $\mathcal{T}_{\mathrm{train}}$ contains the indices of the training snapshots and $\mathcal{A}$ contains the indices of the $n_{a}$ anchor snapshots.

After forming the local neighbourhood $\mathcal{N}_{\mathrm{train}}$ within the training subset without information from the validation subset, the validation points are assigned to the neighbourhood via an adaptive $k$-NN approach. This is done to achieve a comparable local data point distribution; instead of a fixed number of validation points, a variable number of $k_{\mathrm{valid}}(a)$ points is chosen to form the neighbourhood:

\begin{equation}
    \mathcal{N}_{\mathrm{valid}} = \Bigl\{t \in \mathcal{T}_{\mathrm{valid}}: \bigl\|\stateMatrixTime{a} - \stateMatrixTime{t}\bigr\|_{2} \leq r\bigl(a, \, k_{\mathrm{valid}}(a)\bigr)\Bigr\} \, .
\end{equation}

The number of validation points to be added is defined such that the average \textsc{Euclidean} distance of the validation points in the neighbourhood $\mathcal{N}_{\mathrm{valid}}$ to the anchor point matches that of the training points $\mathcal{N}_{\mathrm{train}}$ to the anchor point:

\begin{equation}
    k_{\mathrm{valid}}(a) = \argmin_{k^{\prime} \in \{1, \, \ldots, \, T_{\mathrm{valid}}\}} \, \biggl| \frac{1}{k^{\prime}} \, \sum_{i=1}^{k^{\prime}} \bigl\|\stateMatrixTime{a} - \stateMatrixTime{t_i}\bigr\|_{2} - r_{\mathcal{N}}\bigl(a, \, k\bigr)\biggr| \, ,
\end{equation}

\noindent where $r_{\mathcal{N}}(a, \, k) = \frac{1}{k} \, \sum_{t \in \mathcal{N}_{\mathrm{train}}} \bigl\|\stateMatrixTime{a} - \stateMatrixTime{t}\bigr\|_{2}$ and the indices $\{t_{1}, \, \ldots, \, t_{T_{\mathrm{valid}}}\} \in \mathcal{T}_{\mathrm{valid}}$ are sorted in ascending distance order to the anchor point. This approach ensures that the local resolution remains comparable across subsets.

The nPOD basis $\boldsymbol{\Phi}_{d_{\mathrm{lat}}, \, \mathcal{N}}$ is obtained from the truncated eigendecomposition of the covariance matrix:

\begin{equation}
\boldsymbol{C}_{\mathcal{N}} = \frac{1}{k} \bigl(\stateMatrix_{\mathrm{train}, \, \mathcal{N}} - \boldsymbol{\mu}_{\mathcal{N}}\boldsymbol{1}^{\top}\bigr)\bigl(\stateMatrix_{\mathrm{train}, \, \mathcal{N}} - \boldsymbol{\mu}_{\mathcal{N}}\boldsymbol{1}^{\top}\bigr)^{\top} \, .
\end{equation}

\noindent Here, the covariance matrix, and in particular, the centre of the neighbourhood $\boldsymbol{\mu}_{\mathcal{N}}$ are based on the local training data:

\begin{equation}
    \stateMatrix_{\mathrm{train}, \, \mathcal{N}} = \bigl[\stateMatrixTime{t}\bigr]_{t \in \mathcal{N}_{\mathrm{train}}} \, .
\end{equation}

On the one hand, the local reconstruction with respect to the nPOD basis is obtained with:

\begin{equation}
    \bigl[\stateMatrixRec_{\mathcal{N}}\bigr]_{\mathrm{nPOD}} = \boldsymbol{\Phi}_{d_{\mathrm{lat}}, \, \mathcal{N}}\boldsymbol{\Phi}_{d_{\mathrm{lat}}, \, \mathcal{N}}^\top \bigl(\stateMatrix_{\mathcal{N}} - \boldsymbol{\mu}_{\mathcal{N}}\boldsymbol{1}^{\top}\bigr) + \boldsymbol{\mu}_{\mathcal{N}}\boldsymbol{1}^{\top}
\end{equation}

via projection and subsequent back-projection. On the other hand, the local reconstruction with respect to the POD basis is defined with:

\begin{equation}
    \bigl[\stateMatrixRec_{\mathcal{N}}\bigr]_{\mathrm{POD}} = \boldsymbol{\Phi}_{d_{\mathrm{lat}}}\boldsymbol{\Phi}_{d_{\mathrm{lat}}}^{\top} \bigl(\stateMatrix_{\mathcal{N}} - \boldsymbol{\mu}_{\mathcal{N}}\boldsymbol{\mathrm{1}}^{\top}\bigr) + \boldsymbol{\mu}_{\mathcal{N}}\boldsymbol{\mathrm{1}}^{\top} \, .
\end{equation}

The difference between the averaged local reconstruction errors of the validation points with respect to both bases is then given as:

\begin{equation}
\begin{split}
    \Delta\overline{e_{\mathrm{NMSE}, \, \mathcal{N}}}(\overline{r}_{\mathcal {N}}, \, d_\mathrm{lat}) = \frac{1}{n_{a}}\sum_{a \in \mathcal{A}} &e_{\mathrm{NMSE}}\bigl(\bigl[\stateMatrixRec_{\mathcal{N}}\bigr]_{\mathrm{POD}}, \, \stateMatrix_{\mathcal{N}}\bigr) \\ -&e_{\mathrm{NMSE}}\bigl(\bigl[\stateMatrixRec_{\mathcal{N}}\bigr]_{\mathrm{nPOD}}, \, \stateMatrix_{\mathcal{N}}\bigr)
\end{split}
\end{equation}

\noindent where the averaging is carried out across neighbourhoods that contain the same number of training points and hence have a comparable neighbourhood radius $r_{\mathcal{N}}$.

The LTR is defined as the maximum of this difference over the neighbourhood radius:

\begin{equation}
    \label{eqn:LTR_definition_2}
    e_{\mathrm{LTR}}(d_{\mathrm{lat}}) = \max_{\overline{r}_{\mathcal{N}}} \, \Delta\overline{e_{\mathrm{NMSE}, \, \mathcal{N}}}(\overline{r}_{\mathcal{N}}, \, d_\mathrm{lat}) \, .
\end{equation}

\FloatBarrier

\section{Estimation of computational costs}
\label{app:computational_cost_estimation}

\paragraph{Proper orthogonal decomposition}
The POD modes are computed using an ARPACK implementation of the \emph{implicitly restarted} \textsc{Lanczos} \emph{method} (IRLM)~\cite{Lehoucq1998}, applied to the spatial covariance matrix $\boldsymbol{C} =\stateMatrix{\stateMatrix}^{\top} \in \mathbb{R}^{3\numberOfGridPoints \times 3\numberOfGridPoints}$. Since $T \gg 3\numberOfGridPoints$ holds for the datasets considered here, the classical POD formulation is used over the method of snapshots. The computation of the covariance matrix dominates the computational cost for this formulation and requires about $T \cdot (3\numberOfGridPoints)^2$ FLOPs, presented as the lower bound in Figure~\ref{fig:computational_cost}a. Instead of estimating the additional cost of the IRLM, an upper bound for the subsequent eigendecomposition is provided in Figure~\ref{fig:computational_cost}a, based on \textsc{Householder} \emph{tridiagonalisation} followed by the \emph{symmetric QR algorithm} with accumulation of the eigenvectors, which together require about $9 \cdot (3\numberOfGridPoints)^3$ FLOPs~\cite{Golub2013}. As the IRLM extracts only the $d_{\mathrm{lat}}$ leading eigenvectors rather than all $3\numberOfGridPoints$, the true cost lies between these two limits and close to the lower
one. In the POD context a dataset forward pass is equivalent to two matrix multiplications (refer to Equations~\ref{eqn:encoder_POD} and~\ref{eqn:decoder_POD}). The cost of this operation is about $4T \cdot (3\numberOfGridPoints) \cdot d_{\mathrm{lat}}$ FLOPs and presented in Figure~\ref{fig:computational_cost}b.

\paragraph{Convolutional autoencoder}
The FLOPs for the CAE are counted using the \emph{fvcore} framework (Meta AI~\cite{Fvcore2026}) and counting is limited to the two main contributors: the convolutional layers and the dense layers. The reported values are doubled to match the FLOP convention introduced above; element-wise operations such as activation functions and normalisation are neglected, as their contribution is small in comparison. As a rough estimate, it is assumed that a single training step consumes three times the computational cost of a forward pass through the network for a single batch, accounting for the forward pass and the backward pass, the latter approximated as twice the forward pass cost~\cite{Kaplan2020}. With $\mathrm{FLOP}_{\mathrm{step}}$ referring to a training step on a single batch, the total reported training cost follows as $\mathrm{FLOP} =
\mathrm{FLOP}_{\mathrm{step}} \cdot n_{\mathrm{batches}} \cdot n_{\mathrm{epochs}}$.

\FloatBarrier

\section{Hyperparameter optimisation}
\label{app:hyperparameter_optimisation}

The hyperparameters defining the CAE architecture and the training configuration were optimised on the wind turbine wake case using Optuna~\cite{Akiba2019} (version 4.1.0) with a TPE sampler. Two separate studies were conducted, one for the yz-plane and one for the xy-plane subset, both at the downstream position $\coordX=3R$.

The search space comprises six hyperparameters governing both the training configuration and the architectural design of the autoencoder. During the studies, the number of latent dimensions was fixed to $d_{\mathrm{lat}} = 16$. Two hyperparameters are continuous and were sampled on a logarithmic scale: the learning rate of the Adam optimiser in the range $[10^{-4},\,10^{-2}]$ and the parameter $\alpha$ of the CELU activation function in the range $[1,\,100]$. The batch size $T_{\mathrm{Batch}}$ was selected from the set $\{32,\,64,\,128,\,256,\,512\}$. The three architectural hyperparameters are categorical, each restricted to three predefined configurations in order to keep the search space tractable.

The first architectural hyperparameter is the number of additional dense layers between the flattened output of the last convolutional stage and the bottleneck layer, with the options $0$, $1$ and $2$. Without additional layers, the flattened feature maps are mapped directly onto the $d_{\mathrm{lat}}$ latent neurons. The $k$-th additional layer before the bottleneck comprises $2^k d_{\mathrm{lat}}$ neurons. For $d_{\mathrm{lat}} = 16$ and two additional layers, the encoder thus maps the flattened feature maps onto $64$, $32$ and finally $16$ neurons. The decoder mirrors this structure and expands the latent representation through $32$ and $64$ neurons.
 
The second architectural hyperparameter defines the number of feature maps $C$ in the convolutional stages, with the decoder again being symmetric to the encoder. Configuration ``F-8'' uses $C=8$ in all convolutional stages. Configuration ``F-16'' increases the number of feature maps to $C=16$ in the last encoder stage, and ``F-32'' further increases it to $C=32$ in the last and $C=16$ in the second-to-last stage, while all preceding stages retain $C=8$.
 
The third architectural hyperparameter specifies the kernel size of the first convolutional layer of the encoder and of the corresponding output layer of the decoder. The options are $3\times3$ (``K-3''), $5\times5$ (``K-5'') and $7\times7$ (``K-7'') kernels.

For each subset, a total of 100 trials was evaluated. To save computational resources, the studies were performed only on the last fold of the six-fold cross-validation described in Section~\ref{sec:methods:dimension_reduction:cae_training}, in which the data were divided into training, validation and test subsets. Each trial was trained for at most 1000 epochs. To prevent overfitting, early stopping was applied: training was terminated when the validation loss had not improved for 100 consecutive epochs, and the model state with the lowest validation loss was retained. The objective of the optimisation was the TKE reconstruction fraction on the validation subset. All decisions within the studies, i.e., the sampling of the TPE, the early stopping and the final selection of the configuration, were based on the validation subset. The test subset was evaluated only after the studies had been completed.

\begin{figure*}[tb]
    \centering
    \includegraphics{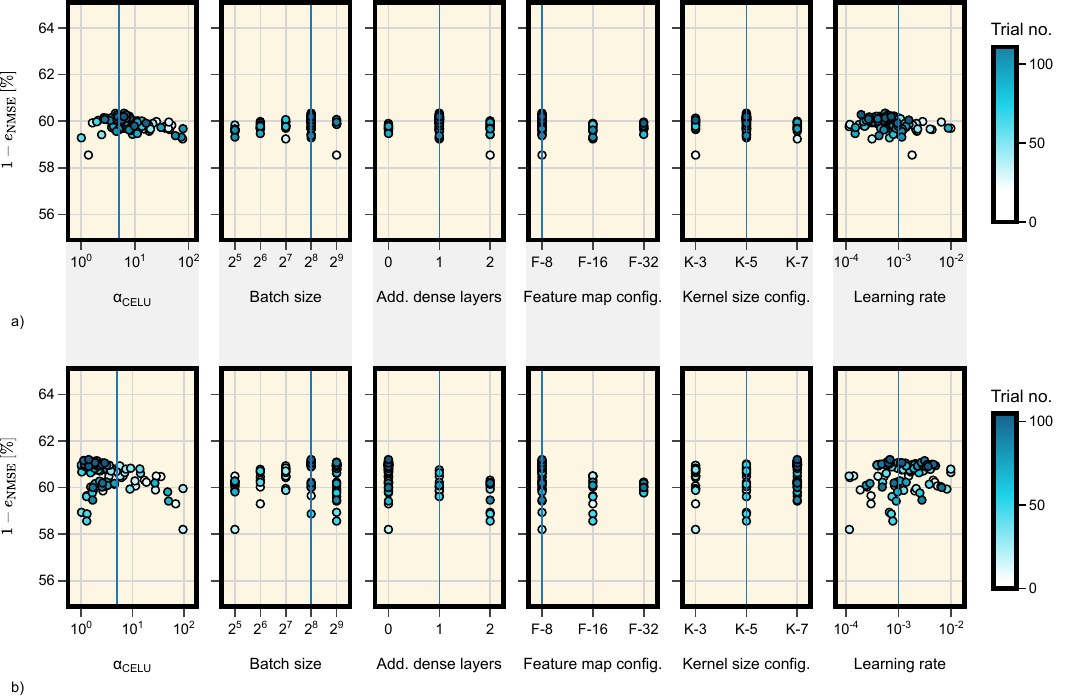}
    \caption{\textbf{Slice plots of the validation TKE reconstruction fraction across the hyperparameter search space.} Each marker represents one trial. The blue line indicates the chosen CAE configuration. \textbf{(a)} yz-plane subset, $\coordX=3R$. \textbf{(b)} xy-plane subset, $\coordX=3R$.}
    \label{fig:hyperparameter_optimisation}
\end{figure*}

The results of both studies are shown in Figure~\ref{fig:hyperparameter_optimisation} as slice plots of the validation TKE reconstruction fraction. Each subplot shows the individual trials as a function of one hyperparameter, which allows the influence of each hyperparameter on model performance to be assessed.

The slice plots reveal an overall narrow spread in the TKE reconstruction fraction across all trials, such that meaningful differences in model performance only become apparent when the objective axis is examined at fine resolution. The $\alpha$ parameter shows the clearest dependence, though its optimal value differs slightly between the two subsets under consideration. For the batch size, the feature map configuration and the learning rate, the best-performing trials of both studies cluster around the same values. The two studies differ in the number of additional dense layers and the kernel size: for the yz-plane, the best results are obtained with one additional dense layer and kernel configuration ``K-5'', whereas for the xy-plane, no additional dense layer and kernel configuration ``K-7'' perform best.

To obtain a single configuration for both subsets, the best configuration of the yz-plane study was adopted. For the xy-plane, the corresponding options, i.e., one additional dense layer and kernel configuration ``K-5'', lie within the spread of the best-performing trials (Figure~\ref{fig:hyperparameter_optimisation}b). This reflects the flat optimisation landscape, in which small variations of the hyperparameters produce no notable change in the TKE reconstruction fraction. The selected configuration is summarised in Table~\ref{tab:hyperparameters}.

\begin{figure}[htb]
    \centering
    \includegraphics{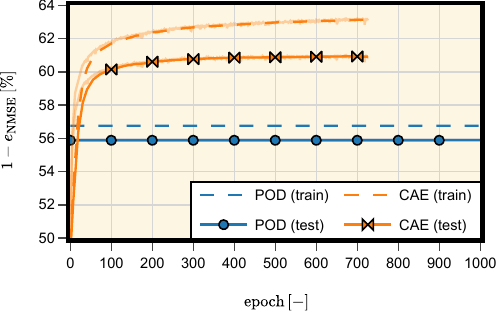}
    \caption{\textbf{Convergence of the TKE reconstruction fraction during CAE training.} Training and test learning curves of the CAE (first fold) over the training epochs for the wind turbine wake case (yz-plane subset, $\coordX=3R$) with $d_{\mathrm{lat}} = 16$ latent dimensions. Faint lines show the raw per-epoch values, bold lines the smoothed values (exponential smoothing). Horizontal lines indicate the POD reference for training and test.}
    \label{fig:convergence_plot}
\end{figure}

Figure~\ref{fig:convergence_plot} shows the learning curves of the selected configuration for the yz-plane subset at $\coordX=3R$ on the first fold. The training was terminated after approximately $725$ epochs. The test TKE reconstruction fraction exceeds the POD reference within the first $20$ epochs and saturates at approximately $61\,\%$ after $200$ to $300$ epochs, which is about $5$ percentage points above POD. While the training reconstruction fraction continues to increase slowly, the test reconstruction fraction does not decrease over the remaining epochs, i.e., no pronounced overfitting occurs. At the end of the training, the gap between training and test is approximately $2$ percentage points, compared with approximately $1$ percentage point for POD.

\begin{figure*}[htb]
    \centering
    \includegraphics{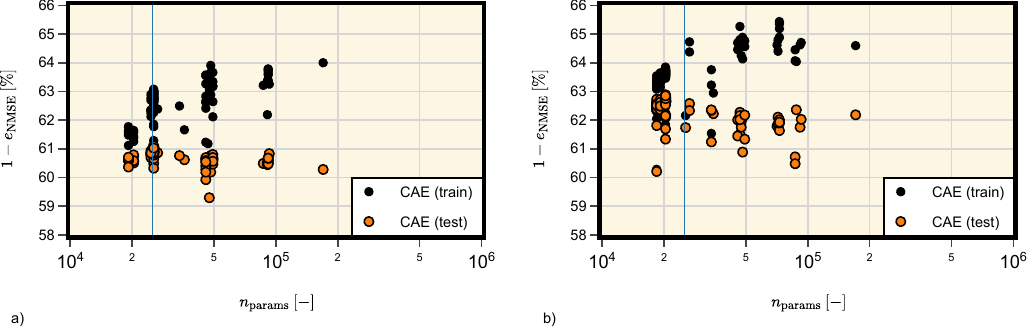} 
    \caption{\textbf{TKE reconstruction fraction over the total number of trainable parameters.} All trials of the two hyperparameter studies are evaluated on the training and test subsets at the early-stopping point. The blue line indicates the chosen CAE configuration. \textbf{(a)} yz-plane subset, $\coordX=3R$. \textbf{(b)} xy-plane subset, $\coordX=3R$.}
    \label{fig:Studies_n_parameters}
\end{figure*}

Figure~\ref{fig:Studies_n_parameters} shows the TKE reconstruction fraction of all trials over the total number of trainable parameters for the training and test subsets at the early-stopping point. The gap between training and test performance increases with the parameter count, indicating increasing overfitting for larger networks. The selected configuration lies in the low-parameter region of the search space (see Table~\ref{tab:trainable_parameters} in Section~\ref{sec:methods:dimension_reduction:cae_architecture_search}). This is consistent with the small gap between the training and test learning curves of the selected configuration in Figure~\ref{fig:convergence_plot}.

\begin{figure*}[htb]
    \centering
    \includegraphics{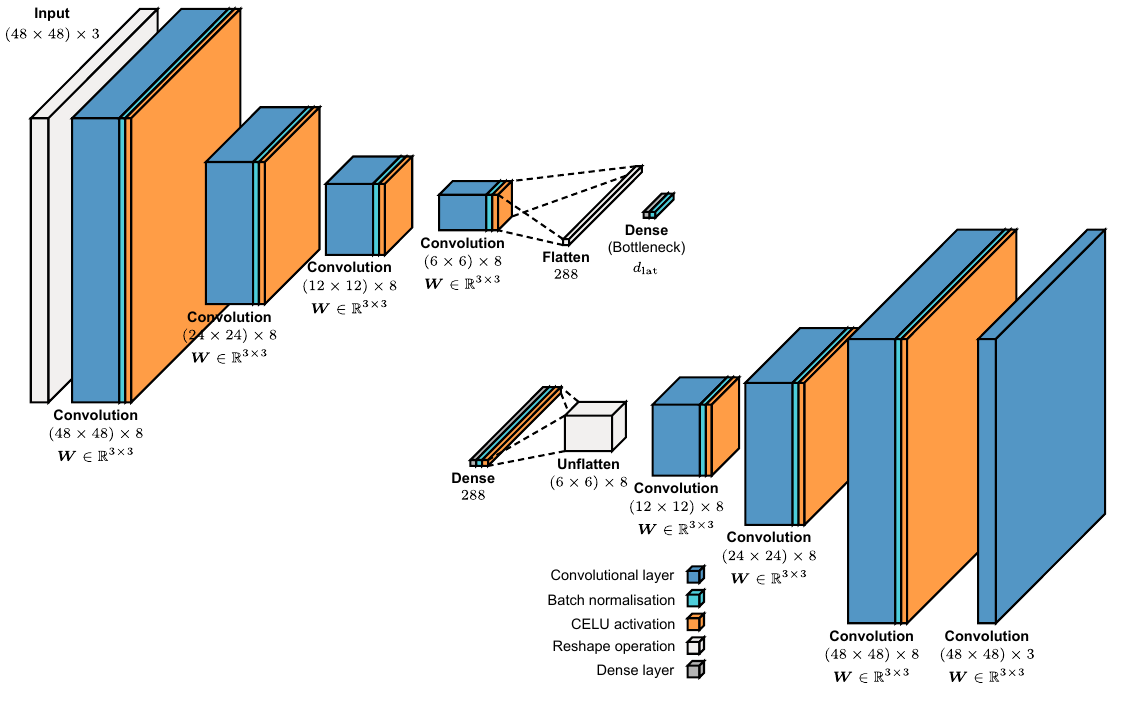}
    \caption{\textbf{Smallest autoencoder architecture in the search space.} All boxes indicate the dimensions of the output of the colour-coded operation. The convolutional units are represented using boxes, where the height and depth of each box represent the spatial dimensions and the width of the box represents the number of feature maps. The dimensions of a feature map are given in parentheses. The figure is not to scale.}
    \label{fig:Autoencoder_small}
\end{figure*}

\begin{figure*}[htb]
    \centering
    \includegraphics{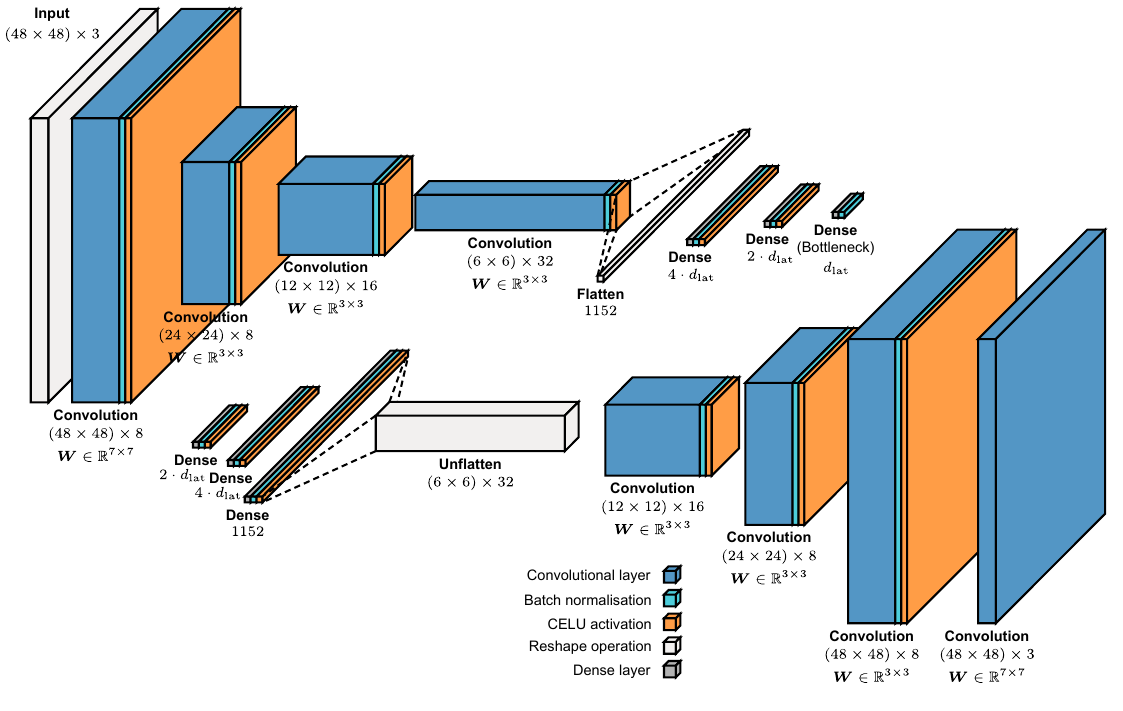}
    \caption{\textbf{Largest autoencoder architecture in the search space.} All boxes indicate the dimensions of the output of the colour-coded operation. The convolutional units are represented using boxes, where the height and depth of each box represent the spatial dimensions and the width of the box represents the number of feature maps. The dimensions of a feature map are given in parentheses. The figure is not to scale.}
    \label{fig:Autoencoder_large}
\end{figure*}

To illustrate the range of architectures covered by the search space, Figures~\ref{fig:Autoencoder_small} and~\ref{fig:Autoencoder_large} sketch the networks with the smallest and the largest possible number of trainable parameters, respectively.

\FloatBarrier

\end{document}